\documentclass[11pt]{article}
\usepackage{amssymb}
\usepackage{makeidx}
\usepackage[totalwidth=460truept,totalheight=612truept]{geometry}
\usepackage{latexsym,amssymb,amsmath,graphicx,accents,eucal,slashed,subfigure,stmaryrd}
\usepackage{dsfont}
\usepackage[T1]{fontenc}
\usepackage[hidelinks]{hyperref}

\global\arraycolsep=1truept

\numberwithin{equation}{section}

\newcommand{\ul}[1]{\mkern2mu\underline{\mkern-2mu #1\mkern-2mu}\mkern2mu }

\begin{document}

\bigskip \phantom{C}

\vskip1.4truecm

\begin{center}
{\huge \textbf{Some Reference Formulas}}

\vskip.5truecm

{\huge \textbf{For The Generating Functions}}

\vskip.5truecm

{\huge \textbf{Of Canonical Transformations}}

\vskip 1truecm

\textsl{Damiano Anselmi}

\vskip .2truecm

\textit{Dipartimento di Fisica ``Enrico Fermi'', Universit\`{a} di Pisa, }

\textit{and INFN, Sezione di Pisa,}

\textit{Largo B. Pontecorvo 3, 56127 Pisa, Italy}

\vskip .2truecm

damiano.anselmi@unipi.it

\vskip 1.5truecm

\textbf{Abstract}
\end{center}

We study some properties of the canonical transformations in classical
mechanics and quantum field theory and give a number of practical formulas
concerning their generating functions. First, we give a diagrammatic formula
for the perturbative expansion of the composition law around the identity
map. Then we propose a standard way to express the generating function of a
canonical transformation by means of a certain \textquotedblleft
componential\textquotedblright\ map, which obeys the
Baker-Campbell-Hausdorff\ formula. We derive the diagrammatic interpretation
of the componential map, work out its relation with the solution of the
Hamilton-Jacobi equation and derive its time-ordered version. Finally, we
generalize the results to the Batalin-Vilkovisky formalism, where the
conjugate variables may have both bosonic and fermionic statistics, and
describe applications to quantum field theory.

\vfill\eject

\section{Introduction}

\label{s0}

\setcounter{equation}{0}

Canonical transformations have a variety of applications, from classical
mechanics to quantum field theory. In particular, they play an important
role when quantum field theory is formulated by means of the functional
integral and the Batalin-Vilkovisky (BV) formalism \cite{bata,weinberg}. The
BV formalism associates external sources $K_{\alpha }$ with the fields $\Phi
^{\alpha }$ and introduces a notion of \textit{antiparentheses} $(X,Y)$ of
functionals $X$, $Y$ of $\Phi $ and $K$. This formal setup is convenient to
treat general gauge theories and study their renormalization, because it
collects the Ward-Takahashi-Slavnov-Taylor (WTST) identities \cite{wtst} in
a compact form and relates in a simple way the identities satisfied by the
classical action $S(\Phi ,K)$ to the identities satisfied by the generating
functional $\Gamma $ of the one-particle irreducible correlation functions. The
canonical transformations, which are the field/source redefinitions that
preserve the antiparentheses, appear in several contexts. For example, they
provide simple ways to gauge fix the theory and map the WTST identities
under arbitrary changes of field variables and gauge fixing. Moreover, they
are a key ingredient of the subtraction of divergences.

The generating functionals of the canonical transformations used in quantum
field theory are often polynomial, and can be composed and inverted with a
small effort. Nevertheless, there are exceptions. When the theory is
nonrenormalizable, for example, as the standard model coupled to quantum
gravity, the canonical transformations involved in the subtraction of the
divergences are nonpolynomial and arbitrarily complicated. Even when the
theory is power counting renormalizable, the variety of fields and sources
that are present and their statistics make it useful to have some shortcuts
and practical formulas to handle the basic operations on canonical
transformations in more straightforward ways.

In this paper, we collect a number of reference formulas concerning the
generating functions of canonical transformations and give diagrammatic
interpretations of their perturbative versions. We first work in classical
mechanics and then generalize the investigation to the BV formalism. The
generalization is actually straightforward, since the operations we define
preserve the statistics of the functionals.

In section \ref{s1} we start from the composition law, by writing the
generating function of the composed canonical transformation as the
tree-level projection of a suitable functional integral. So doing, the
perturbative expansion of the result around the identity map can easily be
expressed in a diagrammatic form. In section \ref{s2} we relate the
composition law to the Baker-Campbell-Hausdorff (BCH) formula \cite{bch}. We
propose a standard way of expressing the generating function of a canonical
transformation by means of a \textit{componential map} $\mathcal{C}(X)$ such
that $\mathcal{C}^{-1}(X)=\mathcal{C}(-X)$ and $\mathcal{C}^{-1}(\mathcal{C}%
(X)\circ \mathcal{C}(Y))=$BCH$(X,Y)$. In section \ref{s3} we derive the
relation between the componential map and the solution of the
Hamilton-Jacobi equation for time-independent Hamiltonians. In section \ref%
{s4} we work out the diagrammatic interpretation of the perturbative
expansion of the componential map around the identity map. In section \ref%
{s5} we generalize the formulas to time-dependent Hamiltonians, which gives
the time-ordered version of the componential map. In section \ref{s6} we
extend the analysis to the BV formalism, where the fields can have arbitrary
statistics. We illustrate a number of applications to quantum field theory.
Section \ref{s7} contains the conclusions.

\section{Composition of canonical transformations}

\label{s1}

\setcounter{equation}{0}

In this section we study the composition of canonical transformations. We
first recall the basic formulas for the generating function of the composite
canonical transformation, in terms of the generating functions of the
components. Then we express the result as the tree-level sector of a
functional integral and provide a diagrammatic interpretation of its
perturbative expansion around the identity map.

Consider two canonical transformations $q_{1},p_{1}\rightarrow q_{2},p_{2}$
and $q_{2},p_{2}\rightarrow q_{3},p_{3}$, with generating functions $%
F_{12}(q_{1},p_{2})$ and $F_{23}(q_{2},p_{3})$, respectively. It is known%
that the generating function
of the composite canonical transformation $q_{1},p_{1}\rightarrow
q_{3},p_{3} $ is%
\begin{equation}
F_{13}(q_{1},p_{3})=F_{12}(q_{1},p_{2})+F_{23}(q_{2},p_{3})-q_{2}^{i}p_{2}^{i},
\label{f13}
\end{equation}%
where $q_{2}^{i}$ and $p_{2}^{i}$ are the functions of $q_{1},p_{3}$ that
extremize the right-hand side.\footnote{%
To our knowledge, very few textbooks report this property. One is ref. \cite%
{guillemin}, where it is ascribed to Hamilton. For a standard derivation,
see also \cite{uwano}. For a derivation from the semiclassical limit of
quantum mechanics, see \cite{davis}. For elaborations from the point of view
of symplectic groupoids, see \cite{cattaneo}.} 

The proof is straightforward. Extremizing the right-hand side with respect
to $q_{2}^{i}$ and $p_{2}^{i}$, we obtain 
\begin{equation*}
0=\frac{\partial F_{12}}{\partial p_{2}^{i}}-q_{2}^{i},\qquad 0=\frac{%
\partial F_{23}}{\partial q_{2}^{i}}-p_{2}^{i}.
\end{equation*}%
Thanks to these equations, the derivatives of $F_{13}$ with respect to $%
q_{1}^{i}$ and $p_{3}^{i}$ can be worked out by keeping $q_{2}^{j}$ and $%
p_{2}^{j}$ constant. This gives the relations%
\begin{equation*}
\frac{\partial F_{13}}{\partial q_{1}^{i}}=\frac{\partial F_{12}}{\partial
q_{1}^{i}}=p_{1}^{i},\qquad \frac{\partial F_{13}}{\partial p_{3}^{i}}=\frac{%
\partial F_{23}}{\partial p_{3}^{i}}=q_{3}^{i},
\end{equation*}%
which prove that $F_{13}(q_{1},p_{3})$ is indeed the generating function of
the canonical transformation $q_{1},p_{1}\rightarrow q_{3},p_{3}$.

We write the composition law as%
\begin{equation}
F_{13}=F_{23}\circ F_{12},  \label{claw}
\end{equation}%
in the sense the $F_{12}$ is the transformation performed first and $F_{23}$
is the one performed last. In particular, given a scalar function $%
S_{1}(q_{1},p_{1})=$ $S_{2}(q_{2},p_{2})=$ $S_{3}(q_{3},p_{3})$, we write%
\begin{equation*}
S_{2}=F_{12}\circ S_{1},\qquad S_{3}=F_{23}\circ S_{2}=F_{23}\circ
F_{12}\circ S_{1}=F_{13}\circ S_{1}.
\end{equation*}%
These formulas mean $%
S_{2}(q_{2},p_{2})=S_{1}(q_{1}(q_{2},p_{2}),p_{1}(q_{2},p_{2}))$, etc.

If we describe the canonical transformations $q_{1},p_{1}\rightarrow
q_{2},p_{2}$ and $q_{2},p_{2}\rightarrow q_{3},p_{3}$ by means of generating
functions $G_{12}(q_{1},q_{2})$ and $G_{23}(q_{2},q_{3})$, then, following
similar steps, it is easy to prove that the composition is generated by%
\begin{equation}
G_{13}(q_{1},q_{3})=G_{12}(q_{1},q_{2})+G_{23}(q_{2},q_{3}),  \label{noa}
\end{equation}%
where $q_{2}$ is the function of $q_{1},q_{3}$ that extremizes the
right-hand side.

In this paper, we are mostly interested in formulas that may have practical
uses in perturbative quantum field theory. It is more convenient to describe
the canonical transformations $q,p\rightarrow Q,P$ by means of generating
functions of the form $F(q,P)$, rather than $G(q,Q)$, because the former can
easily be expanded around the identity transformation and allow us to
express the composite canonical transformation diagrammatically. It is not
possible to achieve these goals in a simple way with generating functions of
the form $G(q,Q)$.

To study the expansion around the identity map, write the generating
functions $F_{12}$ and $F_{23}$ as 
\begin{equation}
F_{A}(q,P)=q^{i}P^{i}+A(q,P),\qquad F_{B}(q,P)=q^{i}P^{i}+B(q,P),
\label{prime}
\end{equation}%
respectively, and their composition $F_{13}$ as 
\begin{equation}
F_{C}(q,P)=q^{i}P^{i}+C(q,P),\qquad F_{C}=F_{B}\circ F_{A}.  \label{seconde}
\end{equation}%
Below we show that the solution $C(q,P)$ can be written as the tree-level
sector of a zero-dimensional functional integral. Thanks to this, the
diagrams that contribute to it can easily be built, according to the
following rules. ($a$) The diagrams, made of lines and vertices, are
connected and contain no loops. ($b$) The vertices are of two types, denoted
by $u$ and $v$, and can have arbitrary numbers of legs. ($c$) Each line of
the diagram must connect one vertex of type $u$ with one vertex of type $v$.

By definition, we include the diagrams that have no lines, that is to say
the vertex $u$ and the vertex $v$. The number of vertices is called \textit{%
order} of the diagram. The absence of loops implies that a diagram of order $%
n$ contains $n-1$ lines, with $n\geqslant 1$. Note that there are no
external legs.

Denote the diagrams of order $n$ by $G_{n\alpha }$, where $\alpha =1,\cdots
,r_{n}$ is an index that labels\ them. Call $f_{n\alpha }$ the combinatorial
factor of $G_{n\alpha }$, which can be calculated with the usual rules, by
viewing $G_{n\alpha }$ as a Feynman diagram. Associate a function $%
C_{n\alpha }(q,P)$ with $G_{n\alpha }$ by replacing each vertex $u$ with the
function $A(q,P)$, each vertex $v$ with the function $B(q,P)$ and each line
with the operator%
\begin{equation}
\frac{\overleftarrow{\partial }}{\partial q^{i}}\frac{\overrightarrow{%
\partial }}{\partial P^{i}},  \label{propa}
\end{equation}%
where the $P$ derivative acts on the function $A$ attached to the line and
the $q$ derivative acts on the function $B$ attached to the line. We call (%
\ref{propa}) the \textit{propagator}.

Then the formula of the function $C(q,P)$ is

\begin{equation}
C(q,P)=\sum_{n=1}^{\infty }C^{(n)}(q,P),\qquad C^{(n)}(q,P)=\sum_{\alpha
=1}^{r_{n}}f_{n\alpha }C_{n\alpha }(q,P).  \label{masterf}
\end{equation}

To prove this result, consider the auxiliary Lagrangian%
\begin{equation*}
\mathcal{L}(\phi ,\psi ,q,P)=A(q,P+\phi )+B(q+\psi ,P)-\psi \phi
\end{equation*}%
and the zero-dimensional quantum field theory described by $\mathcal{L}$,
where $\phi ^{i}$ are $\psi ^{i}$ are the \textquotedblleft
fields\textquotedblright . We focus on the generating function $W(q,P)$
defined by%
\begin{equation*}
\mathrm{e}^{W(q,P)}=\int [\mathrm{d}\phi \mathrm{d}\psi ]\mathrm{e}^{%
\mathcal{L}(\phi ,\psi ,q,P)}.
\end{equation*}%
The square brackets around the measure mean that we consider this integral
as a functional integral, rather than an ordinary one. In other words, we
view it as a bookkeeping for generating diagrams and making standard
operations on diagrams.

The propagator of this theory is determined by the last term of $\mathcal{L}$%
, that is to say $-\psi \phi $, so it is equal to 1. Applying the standard
Feynman rules, it is easy to check that the diagrams defined above give the
tree sector of $W(q,P)$. Clearly, that sector is equal to the Legendre
transform of $\mathcal{L}(\phi ,\psi ,q,P)$ with respect to $\phi $ and $%
\psi $, calculated in zero. Precisely, setting%
\begin{equation}
0=\frac{\partial \mathcal{L}}{\partial \phi ^{i}}=\frac{\partial A}{\partial
P^{i}}(q,P+\phi )-\psi ^{i},\qquad 0=\frac{\partial \mathcal{L}}{\partial
\psi ^{i}}=\frac{\partial B}{\partial q^{i}}(q+\psi ,P)-\phi ^{i},
\label{dera}
\end{equation}%
and denoting the solutions of these conditions by $\phi _{\ast }(q,P)$, $%
\psi _{\ast }(q,P)$, we find%
\begin{equation}
\mathcal{L}(\phi _{\ast },\psi _{\ast },q,P)=A(q,P+\phi _{\ast })+B(q+\psi
_{\ast },P)-\psi _{\ast }\phi _{\ast }.  \label{cl}
\end{equation}

Now, identify $q$ with $q_{1}$ and $P$ with $p_{3}$. Working out $q_{2}$ and 
$p_{2}$ from the canonical transformations generated by $F_{A}(q_{1},p_{2})$
and $F_{B}(q_{2},p_{3})$, given in (\ref{prime}), it is easy to check that%
\begin{equation}
p_{2}^{i}-p_{3}^{i}=\frac{\partial B}{\partial q_{2}^{i}}(q_{2},p_{3}),%
\qquad q_{2}^{i}-q_{1}^{i}=\frac{\partial A}{\partial p_{2}^{i}}%
(q_{1},p_{2}).  \label{ao}
\end{equation}%
On the other hand, formulas (\ref{dera}) give%
\begin{equation}
\phi _{\ast }^{i}=\frac{\partial B}{\partial q_{1}^{i}}(q_{1}+\psi _{\ast
},p_{3}),\qquad \psi _{\ast }^{i}=\frac{\partial A}{\partial p_{3}^{i}}%
(q_{1},p_{3}+\phi _{\ast }).  \label{aa}
\end{equation}%
Expanding (\ref{ao}) and (\ref{aa}) in powers of $A$ and $B$ and comparing
the two outcomes, we get the equalities%
\begin{equation}
\phi _{\ast }^{i}=p_{2}^{i}-p_{3}^{i},\qquad \psi _{\ast
}^{i}=q_{2}^{i}-q_{1}^{i}.  \label{get}
\end{equation}%
Then using (\ref{f13}), (\ref{prime}) and (\ref{seconde}), formula (\ref{cl}%
) gives%
\begin{equation*}
\mathcal{L}(\phi _{\ast },\psi _{\ast
},q_{1},p_{3})=A(q_{1},p_{2})+B(q_{2},p_{3})-(q_{2}^{i}-q_{1}^{i})(p_{2}^{i}-p_{3}^{i})=C(q_{1},p_{3}).
\end{equation*}

We conclude that $C(q,P)$ coincides with $\mathcal{L}(\phi _{\ast },\psi
_{\ast },q,P)$ and is given by the diagrams listed above, which proves (\ref%
{masterf}). We can write%
\begin{equation}
\mathrm{e}^{C(q,P)}=\int^{\prime }[\mathrm{d}\phi \mathrm{d}\psi ]\mathrm{e}%
^{A(q,P+\phi )+B(q+\psi ,P)-\psi \phi },  \label{fint}
\end{equation}%
where the prime on the integral sign means that only the tree contributions
are kept.

For example, the lowest order diagrams contributing to formula (\ref{masterf}%
) are 
\begin{equation}
\includegraphics[width=12truecm,height=5.1truecm]{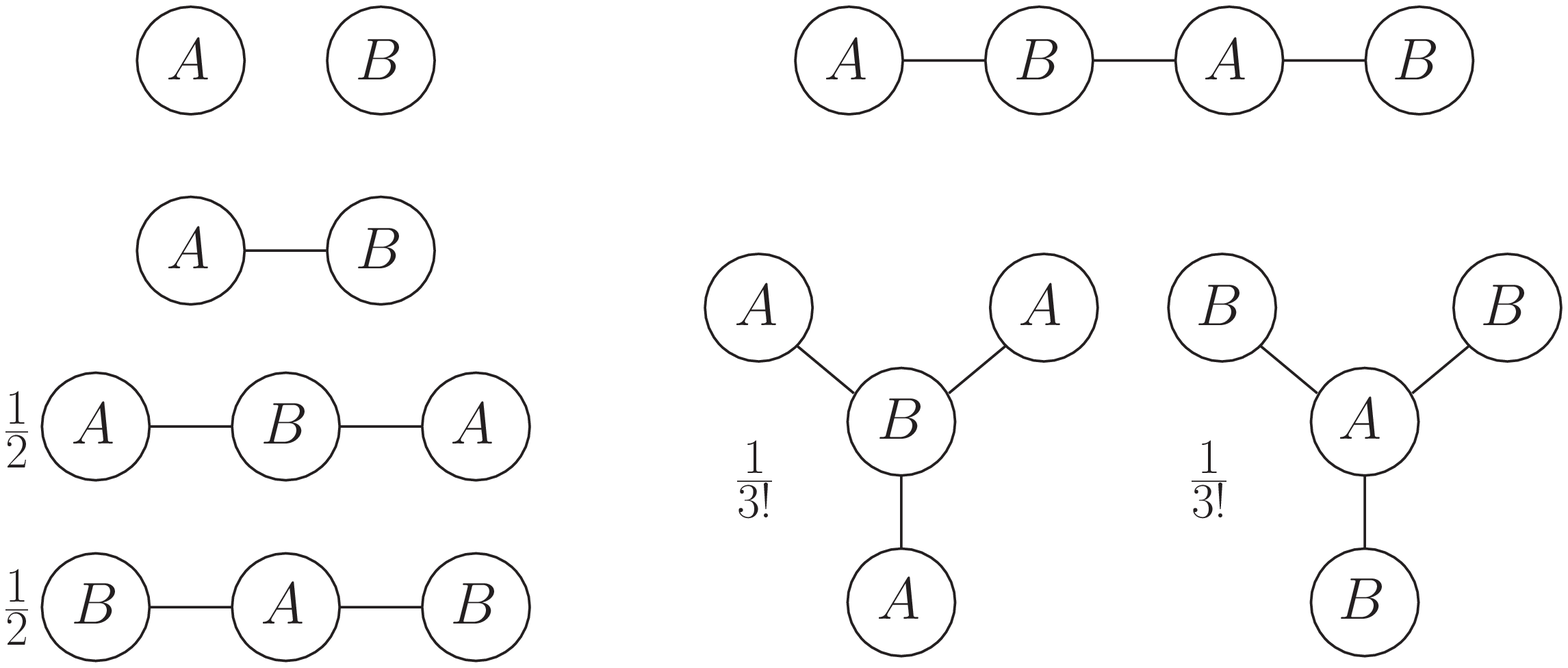}  \label{AB}
\end{equation}%
More explicitly,%
\begin{eqnarray}
C &=&A+B+A_{i}B^{i}+\frac{1}{2}A_{i}B^{ij}A_{j}+\frac{1}{2}B^{i}A_{ij}B^{j}+%
\frac{1}{3!}A_{i}A_{j}A_{k}B^{ijk}+A_{i}B^{ij}A_{jk}B^{k}+\frac{1}{3!}%
B^{i}B^{j}B^{k}A_{ijk}  \notag \\
&&\qquad \qquad +\frac{1}{4!}A_{i}A_{j}A_{k}A_{l}B^{ijkl}+\frac{1}{2}%
A_{i}B^{ij}A_{jk}B^{kl}A_{l}+\frac{1}{2}A_{i}A_{j}A_{kl}B^{ijk}B^{l}  \notag
\\
&&\qquad \qquad +\frac{1}{2}B^{i}B^{j}B^{kl}A_{ijk}A_{l}+\frac{1}{2}%
B^{i}A_{ij}B^{jk}A_{kl}B^{l}+\frac{1}{4!}B^{i}B^{j}B^{k}B^{l}A_{ijkl}+\cdots
,  \label{c2}
\end{eqnarray}%
where%
\begin{equation*}
A_{i_{1}\cdots i_{n}}=\frac{\partial ^{n}A(q,P)}{\partial P_{i_{1}}\cdots
\partial P_{i_{n}}},\qquad B^{i_{1}\cdots i_{n}}=\frac{\partial ^{n}B(q,P)}{%
\partial q_{i_{1}}\cdots \partial q_{i_{n}}}.
\end{equation*}

A simple case is when $A(q,P)=u(q)+f^{i}(q)P^{i}$ for some functions $u(q)$
and $f^{i}(q)$. Then the diagrams give a Taylor expansion that can easily be
resummed into%
\begin{equation}
C(q,P)=A(q,P)+B(q^{i}+f^{i}(q),P).  \label{linear}
\end{equation}%
Similarly, $B(q,P)=v(P)+q^{i}g^{i}(P)$ gives%
\begin{equation}
C(q,P)=A(q,P^{i}+g^{i}(P))+B(q,P).  \label{linear2}
\end{equation}

Another simple case is when $B(q,P)=wB^{\prime }(q,P)$, where $w$ is a
constant parameter that squares to zero, to make the first order of the
Taylor expansion exact. For example, we can take $w=\varpi \varpi ^{\prime }$%
, where $\varpi $ and $\varpi ^{\prime }$ are constant and anticommuting. We
find%
\begin{equation*}
C(q,P)=A(q,P)+B\left( q+\frac{\partial A}{\partial P},P\right) .
\end{equation*}%
Similarly, if $A(q,P)=wA^{\prime }(q,P)$ we have%
\begin{equation*}
C(q,P)=A\left( q,P+\frac{\partial B}{\partial q}\right) +B(q,P).
\end{equation*}

One may wonder if there is a relation between the composition formula (\ref%
{masterf}) and the Baker-Campbell-Hausdorff formula. It turns out that the
formula (\ref{masterf}) is a sort of \textquotedblleft
primitive\textquotedblright\ of the BCH formula. The next section better
clarifies this concept.

\section{The componential map}

\setcounter{equation}{0}

\label{s2}

The composition law of the previous section is good for a number of
purposes, but not practical in other cases. For example, it does not provide
a simple way to invert a canonical transformation. In this section, we
propose a standard way of expressing the generating function of a canonical
transformation by means of a \textquotedblleft
componential\textquotedblright\ map and rephrase the composition law in a
way that makes various properties more apparent. The componential map is
written as a perturbative expansion around the identity map and obeys the
BCH\ formula. Among other things, it makes the inverse operation 
straightforward.

Let $\mathcal{A}$ denote the space of $C^{\infty }$ functions $X,Y,\ldots $
on phase space. Let $\{X,Y\}$ denote the Poisson brackets of $X$ and $Y$,
and $\mathrm{ad}(X):\mathcal{A}\rightarrow \mathcal{A}$, $Y\mapsto \mathrm{ad%
}(X)Y=\{X,Y\}$ denote the adjoint map. Write the BCH formula as 
\begin{equation}
\mathrm{e}^{\mathrm{ad}(X)}\mathrm{e}^{\mathrm{ad}(Y)}=\mathrm{e}^{\mathrm{ad%
}(X+Y+X\triangle Y)},  \label{bosoes}
\end{equation}%
where 
\begin{equation*}
X\triangle Y\equiv \frac{1}{2}\{X,Y\}+\frac{1}{12}\left(
\{X,\{X,Y\}\}+\{Y,\{Y,X\}\}\right) +\cdots
\end{equation*}

The composition law (\ref{claw}) of the previous section defines a map%
\begin{equation*}
\circ \ :\mathcal{A}\times \mathcal{A}\rightarrow \mathcal{A},\qquad \qquad
F_{12},F_{23}\longmapsto F_{13}=F_{23}\circ F_{12}.
\end{equation*}%
The componential map is a map $\mathcal{C}:\mathcal{A}\rightarrow \mathcal{A}
$, $X\longmapsto \mathcal{C}(X)$, such that $\mathcal{C}(0)=I$ and%
\begin{equation}
\mathcal{C}(X)\circ \mathcal{C}(Y)=\mathcal{C}(X+Y+X\triangle Y).
\label{boso}
\end{equation}%
We call it componential map, because it is determined by the composition
law, as we prove below. Note that (\ref{boso}) implies that the
inverse of $\mathcal{C}(X)$ is just $\mathcal{C}(-X)$.

Basically, we regard (\ref{boso}) as an equation for the unknown $\mathcal{C}
$. To better appreciate what we are doing, consider%
\begin{equation*}
E(M)E(N)=E(M+N+M\times N)
\end{equation*}%
as an equation for the unknown map $E$, where $M$ and $N$ are square
matrices of some order, the left-hand side is the matrix product of $E(M)$
and $E(N)$ and $M\times N$ is the same as $M\triangle N$ with Poisson
brackets replaced by commutators. We know that the solution of this problem
is the exponential of the matrix, i.e. $E(M)=\mathrm{e}^{M}$. The
exponential map $\mathrm{e}^{\mathrm{ad}(X)}$ can also be seen as the
solution $\mathcal{E}(X)$ of the equation%
\begin{equation}
\mathcal{E}(X)\mathcal{E}(Y)=\mathcal{E}(X+Y+X\triangle Y),  \label{bosoeq}
\end{equation}%
where $\mathcal{E}(X)$ and $\mathcal{E}(Y)$ are operators $\mathcal{A}%
\rightarrow \mathcal{A}$, and the left-hand side is their product.
Similarly, the componential map is the solution of (\ref{bosoeq}) if $%
\mathcal{E}(X)$ and $\mathcal{E}(Y)$ are viewed as the generating functions
of some canonical transformations and the right-hand side is the generating
function of their composition.

We expand $\mathcal{C}(X)$ as%
\begin{equation}
\mathcal{C}(X)=I+c(X)=I+\sum_{n=1}^{\infty }c_{n}(X),  \label{expa}
\end{equation}%
where $I$ denotes the identity map, $c_{1}=X$ and $c_{n}(X)$, $n\geqslant 2$%
, are homogeneous functions of degree $n$ in $X$ and its derivatives. When
we need to make the arguments of the various functions explicit, we denote
them by $q,P$. Then $I(q,P)=q^{i}P^{i}$ is the generating function of the
identity canonical transformation, while the functions $X$, $\mathcal{C}(X)$%
, $c(X)$, $c_{n}(X)$ are written as $X(q,P)$, $\mathcal{C}(X(q,P))$, $%
c(X(q,P))$ and $c_{n}(X(q,P))$, respectively. Note that the Poisson brackets
involved in the $\triangle $ operation of formula (\ref{boso}) are
calculated with respect to the \textquotedblleft mixed\textquotedblright\
variables $q,P$.

Now we prove that the functions $c_{n}(X(q,P))$, $n>1$, are recursively
determined by the formula%
\begin{equation}
c_{n}(X(q,P))=\frac{1}{n!}\left. \frac{\mathrm{d}^{n-1}}{\mathrm{d}\xi ^{n-1}%
}X\left( q^{i},P^{j}+\sum_{k=1}^{n-1}\xi ^{k}\frac{\partial }{\partial q^{j}}%
c_{k}(X(q,P))\right) \right\vert _{\xi =0}.  \label{thesi}
\end{equation}%
To achieve this goal, we apply the composition law (\ref{boso}) in the
particular case where $X$ and $Y$ are proportional to each other, so that $%
X\triangle Y=0$. If $\sigma $ and $\tau $ are arbitrary constants, we have $%
\mathcal{C}(\sigma X)\circ \mathcal{C}(\tau X)=\mathcal{C}((\sigma +\tau )X)$%
. From formulas (\ref{cl}) and (\ref{expa}), we get%
\begin{equation*}
\sum_{n=1}^{\infty }(\sigma +\tau )^{n}c_{n}(X(q,P))=\sum_{n=1}^{\infty } 
\left[ \tau ^{n}c_{n}(X(q,P+\phi ))+\sigma ^{n}c_{n}(X(q+\psi ,P))\right]
-\psi ^{i}\phi ^{i},
\end{equation*}%
upon extremization with respect to $\phi $ and $\psi $. We differentiate
this equation with respect to $\tau $ and then set $\tau =0$. Because of the
extremization, we can keep $\phi $ and $\psi $ constant. The result is%
\begin{equation}
\sum_{n=1}^{\infty }n\sigma ^{n-1}c_{n}(X(q,P))=X\left(
q,P^{i}+\sum_{n=1}^{\infty }\sigma ^{n}\frac{\partial }{\partial q^{i}}%
c_{n}(X(q,P))\right) ,  \label{ntime}
\end{equation}%
having noted that%
\begin{equation*}
\phi ^{i}=\sum_{n=1}^{\infty }\sigma ^{n}\frac{\partial }{\partial q^{i}}%
c_{n}(X(q,P)),\qquad \psi ^{i}=0,
\end{equation*}%
at $\tau =0$. Differentiating formula (\ref{ntime}) $n-1$ times with respect
to $\sigma $ and setting $\sigma =0$ later on, we get (\ref{thesi}).

The first orders are 
\begin{eqnarray}
&&\mathcal{C}(X)=I+X+\frac{1}{2}X_{i}X^{i}+\frac{1}{3!}\left(
X_{ij}X^{i}X^{j}+X^{j}X_{j}^{i}X_{i}+X^{ij}X_{i}X_{j}\right)  \label{laffo}
\\
&&\quad +\frac{1}{4!}\left(
X_{i}X_{j}^{i}X_{k}^{j}X^{k}+3X_{i}X_{j}^{i}X^{jk}X_{k}+3X^{i}X_{ij}X_{k}^{j}X^{k}+5X^{i}X_{ij}X^{jk}X_{k}\right)
\notag \\
&&\quad +\frac{1}{4!}\left(
X_{ijk}X^{i}X^{j}X^{k}+X_{i}X_{jk}^{i}X^{j}X^{k}+X_{i}X_{j}X_{k}^{ij}X^{k}+X_{i}X_{j}X_{k}X^{ijk}\right) +\cdots ,
\notag
\end{eqnarray}%
where%
\begin{equation*}
X_{j_{i}\cdots j_{m}}^{i_{1}\cdots i_{n}}\equiv \frac{\partial ^{n+m}X(q,P)}{%
\partial q^{i_{1}}\cdots \partial q^{i_{n}}\partial P^{j_{1}}\cdots \partial
P^{j_{m}}}.
\end{equation*}

\section{Relation with the solution of the Hamilton-Jacobi equation}

\setcounter{equation}{0}

\label{s3}

As promised, the componential map is uniquely determined by the composition
law. However, we still have to prove that formula (\ref{boso}) holds for
arbitrary $X$ and $Y$. This goal can be achieved by working out the relation
between the componential map and the solution of the Hamilton-Jacobi
equation.

Rescale $X$ by a factor $\eta $. Recalling that the function $c_{n}$ is
homogeneous of degree $n$, formulas (\ref{expa}) and (\ref{thesi}) give%
\begin{equation*}
\mathcal{C}(\eta X(q,P))=q^{i}P^{i}+\sum_{n=1}^{\infty }\eta
^{n}c_{n}(X(q,P))=q^{i}P^{i}+\sum_{n=1}^{\infty }\frac{\eta ^{n}}{n!}\left.\frac{%
\mathrm{d}^{n-1}}{\mathrm{d}\xi ^{n-1}}X\left( q^{i},\frac{\partial }{%
\partial q^{j}}\mathcal{C}(\xi X(q,P))\right) \right|_{\xi =0}.
\end{equation*}%
This is just the solution of the Hamilton-Jacobi equation%
\begin{equation}
\frac{\partial }{\partial \eta }\mathcal{C}(\eta X(q,P))=X\left( q^{i},\frac{%
\partial }{\partial q^{j}}\mathcal{C}(\eta X(q,P))\right)  \label{hjeq}
\end{equation}%
with the initial condition $\mathcal{C}(0)=I$. To map formula (\ref{hjeq})
into the usual form of the Hamilton-Jacobi equation, it is sufficient to
imagine that $\eta $ is minus the time $t$, the function $X(q,p)$ is a
(time-independent) Hamiltonian $H(q,p)$ and the componential map $\mathcal{C}
$ is the action $S$:%
\begin{equation*}
\frac{\partial S}{\partial t}+H\left( q,\frac{\partial S}{\partial q}\right)
=0.
\end{equation*}

Conversely, given a mechanical system described by the time-independent
Hamiltonian $H(q,p)$, the function 
\begin{equation}
\mathcal{C}(-tH(q,P))=q^{i}P^{i}+\sum_{n=1}^{\infty }(-t)^{n}c_{n}(H(q,P))
\label{sle}
\end{equation}%
is the generating function of the canonical transformation that performs the
time evolution from time $t$ to time zero.

The corresponding Hamilton equations%
\begin{equation}
\frac{\mathrm{d}p^{i}}{\mathrm{d}t}=-\{H(q,p),p^{i}\}=-\mathrm{ad}%
(H(q,p))p^{i},\qquad \frac{\mathrm{d}q^{i}}{\mathrm{d}t}=-\{H(q,p),q^{i}\}=-%
\mathrm{ad}(H(q,p))q^{i},  \label{heq}
\end{equation}%
are solved by the exponential map%
\begin{equation}
Q^{i}=\mathrm{e}^{t\mathrm{ad}(H(q,p))}q^{i},\qquad P^{i}=\mathrm{e}^{t%
\mathrm{ad}(H(q,p))}p^{i}.  \label{0}
\end{equation}%
Indeed, the solution (\ref{sle}) of the Hamilton-Jacobi equation is the
generating function of the canonical transformation that maps $%
q^{i}(t),p^{i}(t)$ to the initial conditions $Q^{i},P^{i}$, because it makes
the transformed Hamiltonian vanish. Clearly, (\ref{heq}) and (\ref{0}) imply 
$\mathrm{d}Q^{i}/\mathrm{d}t=\mathrm{d}P^{i}/\mathrm{d}t=0$. For future
reference, we recall that the Hamilton equations imply%
\begin{equation}
f(Q,P)=\mathrm{e}^{t\mathrm{ad}(H(q,p))}f(q,p),  \label{ham}
\end{equation}%
for an arbitrary function $f\in \mathcal{A}$. Indeed, (\ref{ham}) solves $%
\mathrm{d}f(Q,P)/\mathrm{d}t=0$ and is obviously correct at $t=0$.

Thus, the transformations generated by $\mathcal{C}(X(q,P))$ are 
\begin{equation}
\left( 
\begin{tabular}{l}
$Q^{i}$ \\ 
$P^{i}$%
\end{tabular}%
\right) =\mathrm{e}^{-\mathrm{ad}(X(q,p))}\left( 
\begin{tabular}{l}
$q^{i}$ \\ 
$p^{i}$%
\end{tabular}%
\right) .  \label{1}
\end{equation}%
Since the exponential map satisfies the BCH\ formula (\ref{bosoes}), we can
easily prove that the componential map satisfies the BCH formula (\ref{boso}), 
for arbitrary functions $X$ and $Y$.

To see this, let us write the transformations generated by $\mathcal{C}%
(Y(q_{1},p_{2}))$ and $\mathcal{C}(X(q_{2},p_{3}))$:%
\begin{equation}
\left( 
\begin{tabular}{l}
$q_{3}^{i}$ \\ 
$p_{3}^{i}$%
\end{tabular}%
\right) =\mathrm{e}^{-\mathrm{ad}(X(q_{2},p_{2}))}\left( 
\begin{tabular}{l}
$q_{2}^{i}$ \\ 
$p_{2}^{i}$%
\end{tabular}%
\right) ,\qquad \left( 
\begin{tabular}{l}
$q_{2}^{i}$ \\ 
$p_{2}^{i}$%
\end{tabular}%
\right) =\mathrm{e}^{-\mathrm{ad}(Y(q_{1},p_{1}))}\left( 
\begin{tabular}{l}
$q_{1}^{i}$ \\ 
$p_{1}^{i}$%
\end{tabular}%
\right) .  \label{2}
\end{equation}%
Because of (\ref{claw}), the transformations due to $(\mathcal{C}(X)\circ 
\mathcal{C}(Y))(q_{1},p_{3})$ are then 
\begin{equation}
\left( 
\begin{tabular}{l}
$q_{3}^{i}$ \\ 
$p_{3}^{i}$%
\end{tabular}%
\right) =\mathrm{e}^{-\mathrm{ad}(X(q_{2},p_{2}))}\mathrm{e}^{-\mathrm{ad}%
(Y(q_{1},p_{1}))}\left( 
\begin{tabular}{l}
$q_{1}^{i}$ \\ 
$p_{1}^{i}$%
\end{tabular}%
\right) .  \label{3}
\end{equation}%
Note that the functions $X$ and $Y$ have different arguments in this
formula. To finalize the composition, we must convert $q_{2},p_{2}$ into $%
q_{1},p_{1}$ inside $X(q_{2},p_{2})$. Obviously, the variables used to
calculate the Poisson brackets do not need to be specified, because the
transformations are canonical. In particular, we do not need to specify the
variables in the brackets of the adjoint maps. However, the arguments of $X$
and $Y$ are crucial, which is why we have written them explicitly starting
from formula (\ref{1}).

We have 
\begin{equation*}
X(q_{2},p_{2})=\mathrm{e}^{-\mathrm{ad}(Y(q_{1},p_{1}))}X(q_{1},p_{1}),%
\qquad \mathrm{e}^{-\mathrm{ad}(X(q_{2},p_{2}))}=\mathrm{e}^{-\mathrm{ad}%
(Y(q_{1},p_{1}))}\mathrm{e}^{-\mathrm{ad}(X(q_{1},p_{1}))}\mathrm{e}^{%
\mathrm{ad}(Y(q_{1},p_{1}))}.
\end{equation*}%
The first relation is a particular case of (\ref{ham}), while the second
relation follows from the first one and%
\begin{equation*}
\mathrm{e}^{-\mathrm{ad}(Y)}\{f,g\}=\{\mathrm{e}^{-\mathrm{ad}(Y)}f,\mathrm{e%
}^{-\mathrm{ad}(Y)}g\},
\end{equation*}%
which is another consequence of (\ref{ham}). Then the transformations (\ref%
{3}) become%
\begin{equation*}
\left( 
\begin{tabular}{l}
$q_{3}^{i}$ \\ 
$p_{3}^{i}$%
\end{tabular}%
\right) =\mathrm{e}^{-\mathrm{ad}(Y(q_{1},p_{1}))}\mathrm{e}^{-\mathrm{ad}%
(X(q_{1},p_{1}))}\left( 
\begin{tabular}{l}
$q_{1}^{i}$ \\ 
$p_{1}^{i}$%
\end{tabular}%
\right) .
\end{equation*}%
Since an equivalent version of (\ref{bosoes}) is $\mathrm{e}^{-\mathrm{ad}%
(Y)}\mathrm{e}^{-\mathrm{ad}(X)}=\mathrm{e}^{-\mathrm{ad}(X+Y+X\triangle Y)}$%
, the BCH formula (\ref{boso}) follows by comparison with (\ref{1}) again.

Setting $\mathcal{C}(Y)=F_{A}$, $\mathcal{C}(X)=F_{B}$ and $F_{C}=\mathcal{C}%
(X)\circ \mathcal{C}(Y)$, we can easily check the first few orders of (\ref%
{boso}) by comparing the formulas (\ref{c2}) and (\ref{laffo}).

Summarizing, the componential map is a sort of generating function for the
exponential map. Indeed, the transformations of the coordinates and the
momenta are given by the exponential map and generated by the componential
map.

\section{Diagrammatics of the componential map}

\setcounter{equation}{0}

\label{s4}

We write the diagrammatic expansion of the componential map in the form%
\begin{equation}
\mathcal{C}(X)=I+X+\sum_{n=2}^{\infty }\sum_{G_{nj}\in \mathcal{D}%
_{n}}e_{nj}G_{nj}(X),  \label{main}
\end{equation}%
where $e_{nj}$ are certain coefficients, worked out below, and $\mathcal{D}%
_{n}$ denotes the set of connected tree diagrams $G_{nj}(X)$ built with $n$
vertices $X$ and the propagator (\ref{propa}). Differently from the diagrams
of the previous section, the propagator must carry an arrow, to distinguish
where the $q$ and the $P$ derivatives act. For definiteness, we assume that
the $q$ derivative acts on the $X$ toward which the arrow points and the $P$
derivative acts on the $X$ placed at the other endpoint of the line.

For example, the diagrams of formula (\ref{laffo}) are 
\begin{equation}
\includegraphics[width=16truecm,height=5truecm]{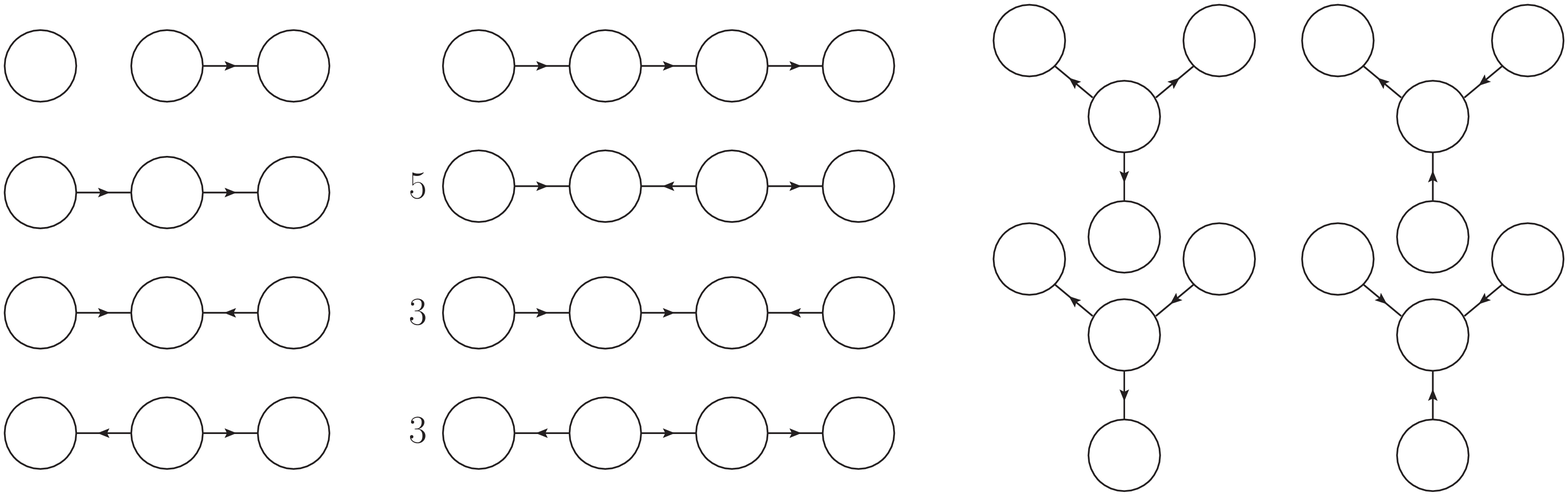}
\label{component}
\end{equation}%
where we have included the coefficients $e_{nj}n!$ different from one. Each
empty disk denotes an $X$.

We work out the rules to calculate the coefficients $e_{nj}$. It is evident
that some of them are simple, others are less straightforward, such as the
factor 5 appearing in the second line of formula (\ref{laffo}). It is
convenient to refer to formula (\ref{thesi}), which gives for $n>1$,%
\begin{equation}
c_{n}(X(q,P))=\frac{1}{n}\sum_{m=1}^{n-1}\sum_{\substack{ \{j_{k}\},\
j_{k}\geqslant 1  \\ j_{1}+\cdots +j_{m}=n-1}}\sigma
_{\{j_{k}\}}X_{i_{1}\cdots i_{m}}(q,P)\prod\limits_{k=1}^{m}\frac{\partial
c_{j_{k}}(X(q,P))}{\partial q^{i_{k}}},  \label{diagra}
\end{equation}%
where the symmetry factor $\sigma _{\{j_{k}\}}$ is equal to one divided by
the product of $\prod\limits_{m}v_{m}!$, $v_{m}$ being the number of times
the integer $m$ appears in the list $\{j_{k}\}$. We recall that $%
c_{1}(X(q,P))=X(q,P)$.

The diagrammatic version of formula (\ref{diagra}) is straightforward,
because the coefficients are just the symmetry factors of the diagrams.
Denote the function $c_{j}$ by means of a disk numbered by $j$. Now the
arrows can only exit $X$ and enter $c_{j}$. For example, we have

\begin{equation*}
\includegraphics[width=16truecm,height=3.5truecm]{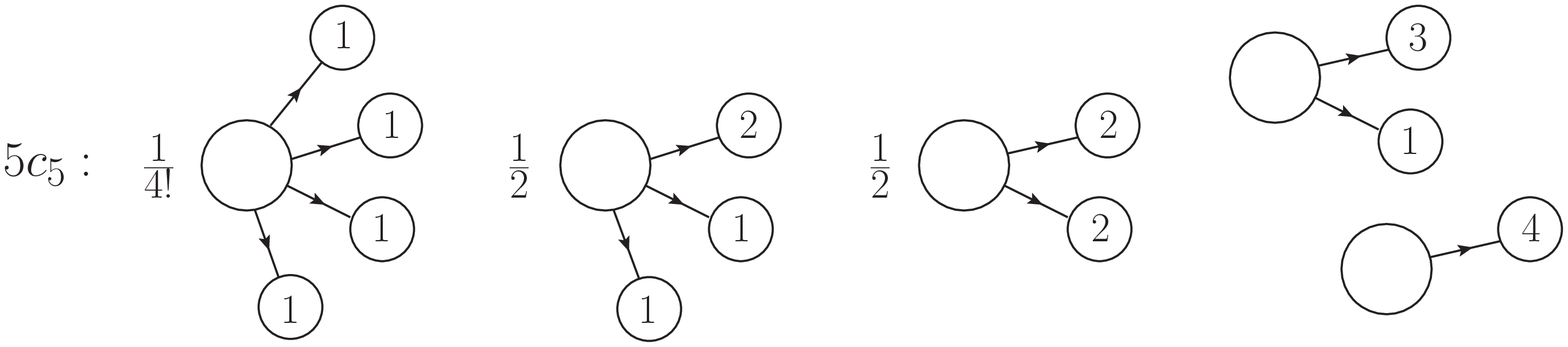}
\end{equation*}%
These diagrammatics generate the diagrammatics of (\ref{main}) by iteration
and allow us to find the rules to compute the coefficients $e_{nj}$. To
formulate these rules, it is useful to define a suitable cutting procedure.

Given a diagram $G_{nj}(X)$, detect the disks to which only exiting lines
are attached. Consider one of such disks at a time. Mark the disk with a
symbol $\times $ at its center and cut the lines attached to the disk in
two. This operation gives a disconnected diagram. For example,

\begin{equation*}
\includegraphics[width=16truecm,height=3truecm]{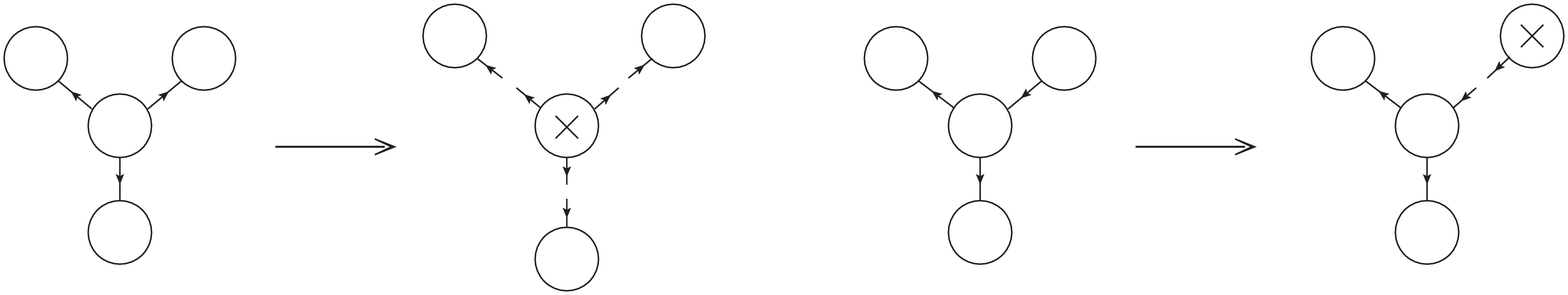}
\end{equation*}%
The so-obtained cut diagrams are made of two types of subdiagrams. One is
the subdiagram made of the marked disk and its lines. The rest is a set of
various subdiagrams $G_{mi}^{\prime }(X)$, each of which is equal to a
diagram of type $G_{mi}(X)$, $m<n$, with one extra incoming line.

To avoid overcounting, coinciding cut diagrams must be counted only once.
For example, the cutting%
\begin{equation*}
\includegraphics[width=11.5truecm,height=1truecm]{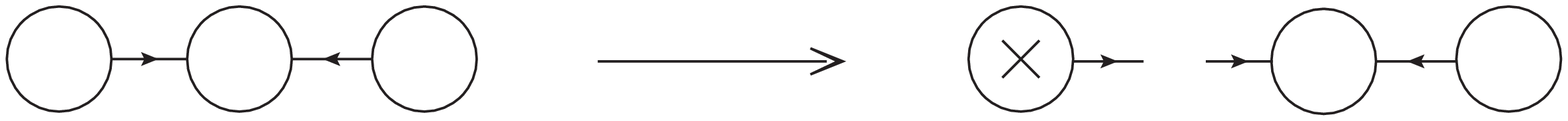}
\end{equation*}%
can be performed in two equivalent ways, by detaching the left disk or the
right one. However, the results are the same, so we must count only one of
them.

Denote the inequivalent cut diagrams by $G_{njk}^{\text{cut}}(X)$, where $k$
is an extra label. Then the coefficient $e_{nj}$ of $G_{nj}$ is given by
the formula%
\begin{equation}
e_{nj}=\frac{1}{n}\sum_{k}e_{njk},  \label{comba}
\end{equation}%
where $e_{njk}$ are coefficients of the cut diagrams $G_{njk}^{\text{cut}}$.
To determine $e_{njk}$,

(i) divide by the number of permutations of the identical subdiagrams $%
G_{mi}^{\prime }$, $m<n$;

(ii) multiply by the number of ways to obtain each subdiagram $%
G_{mi}^{\prime }$, $m<n$, by attaching the extra incoming line to $G_{mi}$;

(iii) multiply by the coefficients $e_{mi}$ of the subdiagrams $G_{mi}$, $%
m<n $.

We illustrate these rules by means of a few examples. First, we see how to
derive the coefficient $5$ of formula (\ref{component}), which corresponds
to $e_{4j}=5/24$. The diagram $G_{4j}$ and its cuts are

\begin{equation*}
\includegraphics[width=16truecm,height=2.5truecm]{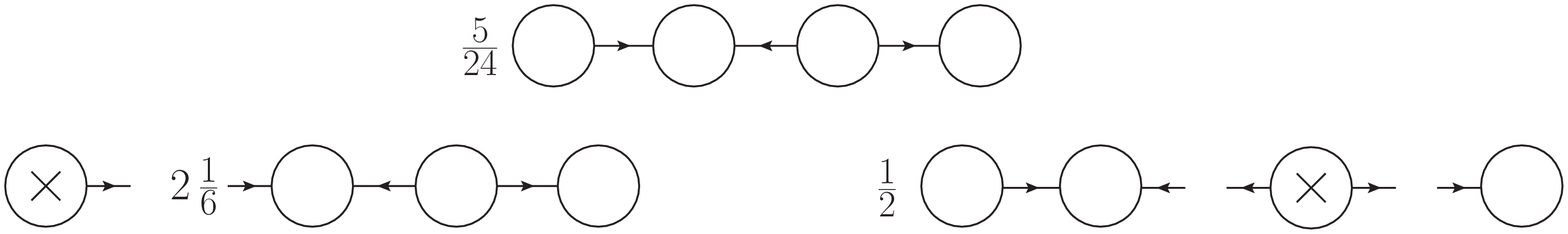}
\end{equation*}%
so we find%
\begin{equation*}
e_{4j}=\frac{1}{4}\left( 2\frac{1}{6}+\frac{1}{2}\right) =\frac{5}{24}.
\end{equation*}%
The reason why the first cut diagram $G_{3i}^{\prime }$ has a factor 2,
besides $e_{3i}=1/6$, is that there are two ways of obtaining $%
G_{3i}^{\prime }$ by attaching the extra incoming line to $G_{3i}$. This is
the meaning of rule (ii).

Next, consider the case

\begin{equation*}
\includegraphics[width=15truecm,height=1.3truecm]{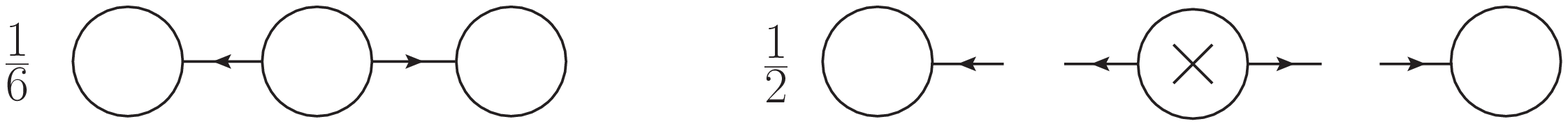}
\end{equation*}%
The factor $1/2$ in front of the cut diagram is due to the permutations of
identical subdiagrams $G_{1i}^{\prime }$. Thus, we have $e_{3i}=1/3(1/2)=1/6$%
. This is the meaning of rule (i).

Formula (\ref{comba}) and the rules just listed are straightforward
consequences of (\ref{diagra}). We have decomposed the diagram $G_{nj}$ into
its contributions as they appear on the right-hand side of (\ref{diagra}),
which are the cut diagrams $G_{njk}^{\text{cut}}$. Each of them has a simple
combinatorial factor $e_{njk}$. The sum of those combinatorial factors,
divided by $n$, gives $e_{nj}$.

An alternative, actually simpler, way to work out the diagrammatic expansion
of the componential map is given in the next section. It follows from the
expansion of the time-ordered componential map, which has straightforward
coefficients. The coefficients of $\mathcal{C}(X)$ are the values of simple
integrals that appear when the time-ordered formula is specialized to the
case of a time-independent function $X$.

Finally, let us mention that we can define the componential logarithm of a
canonical transformation, briefly called c-logarithm, by means of the
inverse componential map. Writing $\mathcal{C}=I+c$ we can invert (\ref%
{laffo}) recursively. The first orders of the c-logarithm are%
\begin{eqnarray}
X &=&c-\frac{1}{2}c_{i}c^{i}+\frac{1}{12}\left(
c_{ij}c^{i}c^{j}+4c^{j}c_{j}^{i}c_{i}+c^{ij}c_{i}c_{j}\right)  \notag \\
&&-\frac{1}{12}\left(
3c_{i}c_{j}^{i}c_{k}^{j}c^{k}+c_{i}c_{j}^{i}c^{jk}c_{k}+c^{i}c_{ij}c_{k}^{j}c^{k}+c^{i}c_{ij}c^{jk}c_{k}+c_{i}c_{jk}^{i}c^{j}c^{k}+c_{i}c_{j}c_{k}^{ij}c^{k}\right) +\cdots
\notag
\end{eqnarray}

\section{Time-ordered componential map}

\setcounter{equation}{0}

\label{s5}

A canonical transformation continuously connected to the identity can be
viewed as a fictitious \textquotedblleft time\textquotedblright\ evolution
associated with a suitable \textquotedblleft Hamiltonian\textquotedblright .
This allows us to relate the componential map to the solution of the
Hamilton-Jacobi equation. In Sect. 4 we have taken advantage of this
correspondence in the case of time-independent Hamiltonians, or, equivalently, 
$\eta $-independent functions $X(q,P)$. Generalizing the formulas of Sect. 4
to time-dependent Hamiltonians $H(q,p,t)$, we can obtain the time-ordered
(precisely, $\eta $-ordered) componential map.

Start from a function $X(q,P,\eta )$ and consider the
Hamilton-Jacobi equation
\begin{equation}
\frac{\partial }{\partial \eta }\mathcal{C}(q,P,\eta )=X\left( q^{i},\frac{%
\partial }{\partial q^{j}}\mathcal{C}(q,P,\eta ),\eta \right) .  \label{xami}
\end{equation}%
Writing $\mathcal{C}(q,P,\eta )=q^{i}P^{i}+c(q,P,\eta )$, we find%
\begin{eqnarray*}
c(q,P,\eta ) &=&\int_{0}^{\eta }\mathrm{d}\eta ^{\prime }X\left( q^{i},P^{j}+%
\frac{\partial }{\partial q^{j}}c(q,P,\eta ^{\prime }),\eta ^{\prime }\right)
\\
&=&\int_{0}^{\eta }\mathrm{d}\eta ^{\prime }X(q,P,\eta ^{\prime
})+\sum_{n=1}^{\infty }\frac{1}{n!}\int_{0}^{\eta }\mathrm{d}\eta ^{\prime
}X_{i_{1}\cdots i_{n}}(q,P,\eta ^{\prime })\prod\limits_{k=1}^{n}\frac{%
\partial c(q,P,\eta ^{\prime })}{\partial q^{i_{k}}},
\end{eqnarray*}%
which can be solved recursively with the help of the following diagrammatics.

Instead of considering the diagrams $G_{nj}$ of the previous section,
consider their $\eta $-ordered versions $\tilde{G}_{nj}$, determined by
applying the following rules. Given a diagram $G_{nj}$, assign coordinates $%
\eta _{k}$ to each disk. We say that

-- the disk with coordinate $\eta _{k}$ is anterior (posterior) to the disk
with coordinate $\eta _{k^{\prime }}$ if $\eta _{k}<\eta _{k^{\prime }}$ ($%
\eta _{k}>\eta _{k^{\prime }}$);

-- a pair of disks is $\eta $-ordered if one of them is anterior to the
other;

-- two disks $D_{1}$ and $D_{2}$ are separated if the path connecting them
(drawn by covering each line only once) contains a third disk $D_{3}$ that
is posterior to both;

-- the latest disk is the one with coordinate $\eta _{k}$ such that $\eta
_{k}>\eta _{k^{\prime }}$ for every $k^{\prime }\neq k$;

-- given a disk $D$, the disk $D^{\prime }$ following $D$ is the most
anterior disk among the disks that are posterior to $D$ and not separated
from $D$.

Assume that the $\eta $ coordinate is the horizontal one and it is oriented
from the right to the left. Displace the disks of $G_{nj}$ so that all the
nonseparated pairs of disks become $\eta $-ordered and each arrow points
from the posterior disk to the anterior one. Two diagrams are said to be
equivalent if every pair of nonseparated disks has the same $\eta $ ordering.

Then, construct all the inequivalent diagrams. Call them $\tilde{G}_{nj}$,
where $n$ is the number of disks and $j$ is an extra label. Denote the set
of diagrams with $n$ disks by $\mathcal{\tilde{D}}_{n}$.

For example, the $\eta $-ordered versions of the diagrams of formula (\ref%
{component}) are 
\begin{equation}
\includegraphics[width=16truecm,height=7truecm]{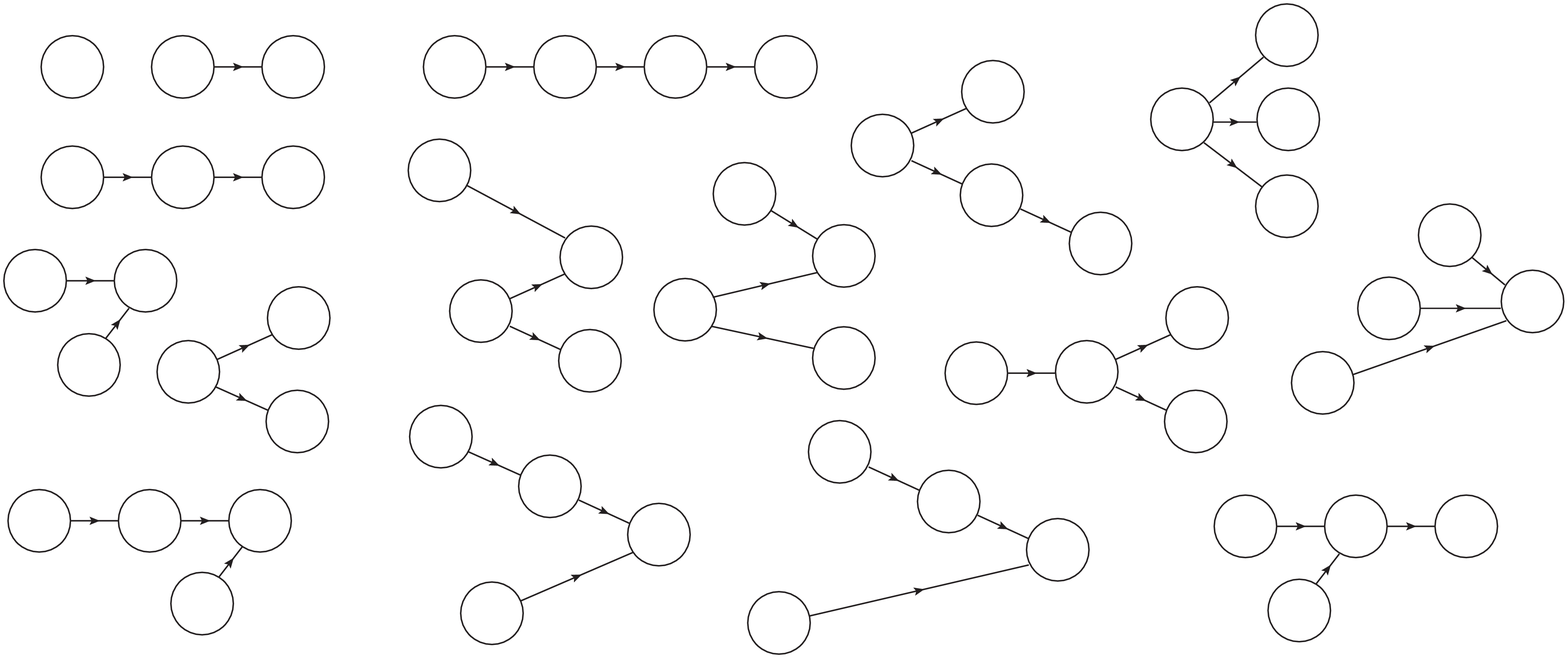}
\label{corde}
\end{equation}

Given a diagram $\tilde{G}_{nj}$, associate a cut diagram $\tilde{G}_{nj}^{%
\text{cut}}$ with it by marking the latest disk with $\times $ and detaching
it from the rest as explained before. The operation generates subdiagrams $%
\tilde{G}_{mj}^{\prime }$, each of which is built by adding an extra
incoming line to a diagram of type $\tilde{G}_{mj}$, with $m<n$. The
symmetry factor of $\tilde{G}_{nj}$ is equal to the product of the symmetry
factors of the subdiagrams $\tilde{G}_{mj}^{\prime }$, divided by the number
of permutations of the equivalent $\tilde{G}_{mj}^{\prime }$s. The symmetry
factor of a subdiagram $\tilde{G}_{mj}^{\prime }$ is equal to the number of
ways to obtain it by adding the extra line to $\tilde{G}_{mj}$, times the
symmetry factor of $\tilde{G}_{mj}$.

Finally, evaluate the diagram $\tilde{G}_{nj}$ as follows. A disk with
coordinate $\eta _{k}$ corresponds to $X(q,P,\eta _{k})$. As before, an
oriented line is the propagator (\ref{propa}), the $q$ derivative acting on
the anterior disk and the $P$ derivative acting on the posterior disk.
Multiply by the symmetry factor of the diagram and integrate the coordinate $%
\eta _{k}$ of each disk from 0 to the coordinate $\eta _{k^{\prime }}$ of
the following disk. Finally integrate the coordinate of the latest disk from
0 to $\eta $. This gives a function $\tilde{G}_{nj}(q,P,\eta )$. The sum of
these functions plus the identity map gives the $\eta $-ordered componential
map, which reads 
\begin{equation}
\mathcal{C}(q,P,\eta )=q^{i}P^{i}+\int_{0}^{\eta }\mathrm{d}\eta ^{\prime
}X(q,P,\eta ^{\prime })+\sum_{n=2}^{\infty }\sum_{\tilde{G}_{nj}\in \mathcal{%
\tilde{D}}_{n}}\tilde{G}_{nj}(q,P,\eta ).  \label{tcomp}
\end{equation}%
To order three we have 
\begin{eqnarray}
\mathcal{C}(q,P,\eta ) &=&q^{i}P^{i}+\int_{0}^{\eta }\mathrm{d}\eta ^{\prime
}X(q,P,\eta ^{\prime })+\int_{0}^{\eta }\mathrm{d}\eta ^{\prime
}X_{i}(q,P,\eta ^{\prime })\int_{0}^{\eta ^{\prime }}\mathrm{d}\eta ^{\prime
\prime }X^{i}(q,P,\eta ^{\prime \prime })  \notag \\
&&+\int_{0}^{\eta }\mathrm{d}\eta ^{\prime }X_{i}(q,P,\eta ^{\prime
})\int_{0}^{\eta ^{\prime }}\mathrm{d}\eta ^{\prime \prime
}X_{j}^{i}(q,P,\eta ^{\prime \prime })\int_{0}^{\eta ^{\prime \prime }}%
\mathrm{d}\eta ^{\prime \prime \prime }X^{j}(q,P,\eta ^{\prime \prime \prime
})  \notag \\
&&+\int_{0}^{\eta }\mathrm{d}\eta ^{\prime }X_{i}(q,P,\eta ^{\prime
})\int_{0}^{\eta ^{\prime }}\mathrm{d}\eta ^{\prime \prime }X_{j}(q,P,\eta
^{\prime \prime })\int_{0}^{\eta ^{\prime \prime }}\mathrm{d}\eta ^{\prime
\prime \prime }X^{ij}(q,P,\eta ^{\prime \prime \prime })  \label{cpn} \\
&&+\frac{1}{2}\int_{0}^{\eta }\mathrm{d}\eta ^{\prime }X_{ij}(q,P,\eta
^{\prime })\int_{0}^{\eta ^{\prime }}\mathrm{d}\eta ^{\prime \prime
}X^{i}(q,P,\eta ^{\prime \prime })\int_{0}^{\eta ^{\prime }}\mathrm{d}\eta
^{\prime \prime \prime }X^{j}(q,P,\eta ^{\prime \prime \prime })+\cdots 
\notag
\end{eqnarray}

As anticipated before, an alternative way to compute the coefficients $%
e_{nj} $ and $e_{njk}$ of formulas (\ref{main}) and (\ref{comba}) is to use
formula (\ref{tcomp}), assume that $X$ is $\eta $ independent, integrate the
various coordinates $\eta _{k}$ and finally set $\eta =1$.\ Diagrams that
are identical for the purposes of the previous section have different $\eta $
orderings, which is why the coefficients of the $\eta $-ordered componential
map are much simpler than $e_{nj}$ and $e_{njk}$.

When we have a one-parameter family of generating functions $\mathcal{C}%
(q,P,\eta )$ such that $\mathcal{C}(q,P,0)=I(q,P)$, we can give a more
practical definition of logarithm. Viewing fictitiously the $\eta $
dependence as a time evolution, we define the h-logarithm (h standing for
\textquotedblleft Hamiltonian\textquotedblright ) as the Hamiltonian $%
X(q,p,\eta )$ associated with it. By the Hamilton-Jacobi equation (\ref{xami}%
), we have 
\begin{equation}
X(q,p,\eta )=\widetilde{\frac{\partial \mathcal{C}}{\partial \eta }},
\label{re}
\end{equation}%
where the tilde means that the argument $P$ must be solved in terms of $%
q,p,\eta $ by means of the canonical transformation $\mathcal{C}$ itself.
For future use we remark that, in particular, if $f(q,p,\eta )$ is a
function that behaves as a scalar under $\mathcal{C}$, i.e. such that $%
f^{\prime }(Q,P,\eta )=f(q,p,\eta )$, we have%
\begin{equation}
\frac{\partial f^{\prime }}{\partial \eta }=\frac{\partial f}{\partial \eta }%
-\left\{ f,\widetilde{\frac{\partial \mathcal{C}}{\partial \eta }}\right\} .
\label{remo}
\end{equation}%
If there is no parameter $\eta $ to apply (\ref{re}), the h-logarithm is not
defined. If $\mathcal{C}(q,P,\eta ,\zeta ,\ldots )$ depends on more
parameters $\eta ,\zeta ,\ldots $ and $\mathcal{C}(q,P,0,0,\ldots )$
coincides with the identity map, we have one h-logarithm for each parameter.
In the time-independent case $\mathcal{C}(\eta X(q,P))$, the h-logarithm $%
X(q,p,\eta )$ coincides with $X(q,p)$. Note that the c-logarithm always
exists and is unique.

\section{Canonical transformations and Batalin-Vilkovisky formalism}

\setcounter{equation}{0}

\label{s6}

In this section we generalize the results found so far to the
Batalin-Vilkovisky formalism, where the generating function(al)s are
fermionic and the fields may be both bosonic and fermionic. Then we give
some examples that have applications to both renormalizable and
nonrenormalizable theories. We compose the canonical transformations that
perform the gauge fixing with those that switch to the background field
method. Then we use the componential map to interpolate between the
background field approach and the standard nonbackground approach.

The Batalin-Vilkovisky formalism is convenient to study general gauge
theories. The conjugate variables are the fields $\Phi ^{\alpha }$ and
certain external sources $K_{\alpha }$ coupled to the $\Phi $ symmetry
transformations. A notion of \textit{antiparentheses}%
\begin{equation}
(X,Y)\equiv \int \left( \frac{\delta _{r}X}{\delta \Phi ^{\alpha }}\frac{%
\delta _{l}Y}{\delta K_{\alpha }}-\frac{\delta _{r}X}{\delta K_{\alpha }}%
\frac{\delta _{l}Y}{\delta \Phi ^{\alpha }}\right)  \label{antip}
\end{equation}%
is introduced, where $X$ and $Y$ are functionals of $\Phi $ and $K$, the
integral is over spacetime points associated with repeated indices and the
subscripts $l$ and $r$ in $\delta _{l}$ and $\delta _{r}$ denote the left
and right functional derivatives, respectively. The fields $\Phi ^{\alpha }$
and the sources $K_{\alpha }$ have statistics $\varepsilon _{\alpha }$ and $%
\varepsilon _{\alpha }+1$, respectively, which are equal to 0 mod 2 for
bosons and 1 mod 2 for fermions.

The fields $\Phi ^{\alpha }$ include the classical fields $\phi ^{i}$, the
Fadeev-Popov ghosts $C^{I}$, the antighosts $\bar{C}^{I}$ and the Lagrange
multipliers $B^{I}$ for the gauge fixing. The action $S(\Phi ,K)$ is a local
functional that satisfies the master equation $(S,S)=0$ and coincides with
the classical action $S_{c}(\phi )$ at $C=\bar{C}=B=K=0$.

The canonical transformations are the transformations $\Phi ,K\rightarrow
\Phi ^{\prime },K^{\prime }$ that preserve the antiparentheses (\ref{antip}%
). They can be derived from a generating functional $F(\Phi ,K^{\prime })$
of fermionic statistics, by means of the formulas%
\begin{equation*}
\Phi ^{\alpha \hspace{0.01in}\prime }=\frac{\delta F}{\delta K_{\alpha
}^{\prime }},\qquad K_{\alpha }=\frac{\delta F}{\delta \Phi ^{\alpha }}.
\end{equation*}%
The identity transformation is generated by $F(\Phi ,K^{\prime })=\int \Phi
^{\alpha }K_{\alpha }^{\prime }$.

The formulas derived in the previous sections for the componential map and
the composition of canonical transformations can be immediately generalized
to fermionic functionals of fields and sources of various statistics.
Indeed, the basic operator, that is to say the propagator (\ref{propa}), is
turned into%
\begin{equation}
\int \frac{\overleftarrow{\delta_{r} }}{\delta \Phi ^{\alpha }(x)}\frac{%
\overrightarrow{\delta_{l} }}{\delta K_{\alpha }^{\prime }(x)},
\label{propa2}
\end{equation}%
which has fermionic statistics. The functionals $F(\Phi ,K^{\prime })$, $%
\mathcal{C}(X)$ and $X$ also have fermionic statistics. Thus, each time we
add a propagator and a new disk $X$, the statistics are correctly preserved.
As a consequence, the formulas found so far can be straightforwardly applied
to the BV formalism.

Canonical transformations are used for various purposes in quantum field
theory. They encode the most general (changes of) gauge fixing and changes
of field variables. Moreover, they are an important ingredient of the
perturbative subtraction of divergences. Precisely, they subtract the
divergences that are proportional to the field equations. The composition
and the inversion of canonical transformations are operations that are met
frequently. Often, it is enough to study them at the infinitesimal level,
but sometimes it is necessary to handle them exactly or to all orders of the
expansion. The literature on these topics is wide, both at the
mathematical/formal level \cite{bata,tyutin} and at the level of
renormalization and gauge dependence \cite%
{voronov,removal,quadri2,recentgaugedep,ABward}.

We recall that the BV\ formalism is quite versatile and can be used to
formulate all kinds of general gauge theories, including those where the
symmetry transformations close only on shell and those that have reducible
gauge algebras (where the ghosts have local gauge symmetries of their own
and it is necessary to introduce \textquotedblleft ghosts of
ghosts\textquotedblright ). Our formulas hold in those cases also.

Nevertheless, we concentrate the applications of this section to the
irreducible gauge symmetries that close off shell, which have the most
important applications to physics. In those cases, there exists a solution $S(\Phi
,K) $ of the master equation that is linear in $K$:%
\begin{equation}
S(\Phi ,K)=S_{c}(\phi )-\int R^{\alpha }(\Phi )K_{\alpha }\text{.}
\label{sfk}
\end{equation}%
The functions $R^{\alpha }(\Phi )$ are the symmetry transformations of the
fields $\Phi ^{\alpha }$. See for example the appendix of ref. \cite{ABward}
for explicit formulas in the case of general covariance, local Lorentz
symmetry, Abelian gauge symmetries and non-Abelian Yang-Mills symmetries.

We give some examples of applications in the context of the background field
method \cite{backf}. Two different approaches to formulate the background
field method in the context of the BV formalism can be found in the
literature, the one of refs. \cite{quadri2,quadri} by Binosi and Quadri%
\footnote{%
See also \cite{grassi} for a similar approach in the language of WTST\
identities and the Zinn-Justin equation.} and the one of the present author 
\cite{back}. The two have properties that are good for different purposes.
Here we follow the approach of \cite{back}. One starts from the action%
\begin{equation}
S(\Phi ,K,\ul{\Phi },%
\ul{K})=S_{c}(\phi )-\int R^{\alpha
}(\Phi )K_{\alpha }-\int R^{\alpha }(\ul{\Phi %
})\ul{K}_{\alpha },  \label{sfb}
\end{equation}%
which is obtained from (\ref{sfk}) by adding a background copy with
vanishing classical action. It is not necessary to have background copies of
the antighosts and the Lagrange multipliers, so we take $\ul{%
\Phi }^{\alpha }=\{\ul{\phi }^{i},\ul{C}%
^{I}\}$ and $\ul{K}_{\alpha }=\{\ul{K}_{\phi
}^{i},\ul{K}_{C}^{I}\}$, where $%
\ul{\phi }^{i}$ and $%
\ul{C}^{I}$ are background copies of the
physical fields and the ghosts, respectively, and $\ul{K}_{\phi }^{i}$, $\ul{K%
}_{C}^{I}$ are the sources associated with them.

Then we perform the background shift, by means of the canonical
transformation generated by\footnote{%
Differently from ref. \cite{back}, we understand that the fields and the
sources with primes are the transformed ones. This originates some sign
differences with respect to the formulas of \cite{back}.}%
\begin{equation*}
F_{\text{b}}(\Phi ,\ul{ \Phi }
,K^{\prime },\ul{ K} ^{\prime
})=\int (\Phi ^{\alpha }-\ul{ \Phi }%
 ^{\alpha })K_{\alpha }^{\prime }+\int \ul{ \Phi } ^{\alpha }\ul{ K%
} _{\alpha }^{\prime }.
\end{equation*}%
Taking advantage of the componential map, we can write%
\begin{equation*}
F_{\text{b}}=\mathcal{C}\left( -\int \ul{ \Phi %
} ^{\alpha }K_{\alpha }^{\prime }\right) .
\end{equation*}%
Indeed, the argument of $\mathcal{C}$ does not depend on any pair of
conjugate variables, so all the nontrivial diagrams of formula (\ref{main})
vanish.

After the shift, the action is $F_{\text{b}}S$. The new fields $\Phi
^{\alpha }$ are called quantum fields. The symmetry transformations $%
R^{i}(\Phi )$ of $\phi ^{i}$ are turned into the transformations $R^{i}(\Phi
+\ul{ \Phi } )$ of $\phi ^{i}+%
\ul{ \phi } ^{i}$. These can be
decomposed as the sum of the background transformations $R^{i}(%
\ul{ \Phi } )$ of $\ul{%
 \phi } ^{i}$ plus the transformations $%
R^{i}(\Phi +\ul{ \Phi } )-R^{i}(%
\ul{ \Phi } )$ of $\phi ^{i}$.
In turn, the transformations of $\phi ^{i}$ split into the sum of the
quantum transformations of $\phi ^{i}$ [made of the $\ul{%
 C} $-independent part of $R^{i}(\Phi +%
\ul{ \Phi } )-R^{i}(\ul{%
 \Phi } )$], plus the background
transformations of $\phi ^{i}$ (the $\ul{ C} $-dependent part). Something similar happens to the symmetry
transformations of the ghosts $C$.

The background transformations of the antighosts and the Lagrange
multipliers remain trivial after $F_{\text{b}}$, and need to be adjusted by
means of a further canonical transformation, generated by%
\begin{equation*}
F_{\text{nm}}(\Phi ,\ul{ \Phi }
,K^{\prime },\ul{ K} ^{\prime
})=\int \Phi ^{\alpha }K_{\alpha }^{\prime }+\int \ul{ \Phi } ^{\alpha }\ul{ K%
} _{\alpha }^{\prime }-\int \mathcal{R}_{\bar{C}}^{I}(%
\bar{C},\ul{ C} )K_{B}^{I%
\hspace{0.01in}\prime }=\mathcal{C}\left( -\int \mathcal{R}_{\bar{C}}^{I}(%
\bar{C},\ul{ C} )K_{B}^{I%
\hspace{0.01in}\prime }\right) ,
\end{equation*}%
where $\mathcal{R}_{\bar{C}}^{I}(\bar{C},\ul{ C%
} )$ denotes the background transformation of the
antighosts. Explicitly, the argument of the componential map $\mathcal{C}$
is 
\begin{equation}
\int (gf^{abc}\ul{ C} ^{b}\bar{C%
}^{c}+\ul{ C} ^{\rho }\partial
_{\rho }\bar{C}^{a})K_{B}^{a\hspace{0.01in}\prime }+\int (2%
\ul{ C} ^{\hat{a}\hat{c}}\eta _{\hat{c}%
\hat{d}}\bar{C}^{\hat{d}\hat{b}}+\ul{ C}%
 ^{\rho }\partial _{\rho }\bar{C}^{\hat{a}\hat{b}})K_{\hat{a}\hat{b}%
B}^{\hspace{0.01in}\prime }+\int \left( \ul{ C%
} ^{\rho }\partial _{\rho }\bar{C}_{\mu }-\bar{C}_{\rho
}\partial _{\mu }\ul{ C} ^{\rho
}\right) K_{B}^{\mu \hspace{0.01in}\prime },  \label{argu}
\end{equation}%
for Yang-Mills symmetries, local Lorentz symmetry and diffeomorphisms, where
the hats on $a,b,\ldots $ are used to distinguish the local Lorentz indices
from the Yang-Mills ones.

Finally, the theory can be gauge fixed in a background invariant way by
means of the canonical transformation generated by%
\begin{equation}
F_{\text{gf}}(\Phi ,\ul{ \Phi }
,K^{\prime },\ul{ K} ^{\prime
})=\int \ \Phi ^{\alpha }K_{\alpha }^{\prime }+\int \ \ul{%
 \Phi } ^{\alpha }\ul{ K} _{\alpha }^{\prime }-\Psi (\Phi ,%
\ul{ \phi } )=\mathcal{C}(-\Psi ),
\label{fgf}
\end{equation}%
where $\Psi (\Phi ,\ul{ \phi }
) $ is a background invariant functional of fermionic statistics, known as
gauge fermion. Typically, we choose it of the form%
\begin{equation*}
\Psi (\Phi ,\ul{ \phi } )=\int 
\bar{C}^{I}\left( G^{Ii}(\ul{ \phi }%
 ,\partial )\phi ^{i}+\zeta _{IJ}(\ul{
\phi } ,\partial )B^{J}\right) ,
\end{equation*}%
where $G^{Ii}(\ul{ \phi }
,\partial )\phi ^{i}$ are the gauge-fixing functions. It is common to choose
such functions to be linear in the quantum fields $\phi ^{i}$, to simplify
various properties of renormalization. The operator matrix $\zeta _{IJ}(%
\ul{ \phi } ,\partial )$ is
symmetric, nonsingular at $\ul{ \phi }%
 =0$ and proportional to the identity in every simple subgroup of
the gauge symmetry group. The relation $F_{\text{gf}}=\mathcal{C}(-\Psi )$
of (\ref{fgf}) follows from the fact that the gauge fermion does not depend
on the sources $K$.

Invariance under background transformations is easy to achieve, by combining
the plain derivative $\partial $ with the background field $%
\ul{ \phi } $ to build the background
covariant derivative. For example, we can take%
\begin{eqnarray*}
\Psi &=&\int \sqrt{|\ul{ g} |}%
\bar{C}^{a}\left( \ul{ g} ^{\mu
\nu }D_{\mu }(\ul{ A} ,%
\ul{ g} )A_{\nu }^{a}+\zeta
_{1}B^{a}\right) , \\
\Psi &=&\int \sqrt{|\ul{ g} |}%
\bar{C}_{\hat{a}\hat{b}}\left( \ul{ e}%
 ^{\rho \hat{a}}\ul{ g} ^{\mu \nu }D_{\mu }(\ul{ e}
)D_{\nu }(\ul{ e} )f_{\rho }^{%
\hat{b}}+\frac{\zeta _{2}}{2}B^{\hat{a}\hat{b}}+\frac{\zeta _{3}}{2}%
\ul{ g} ^{\mu \nu }D_{\mu }(%
\ul{ e} )D_{\nu }(\ul{%
 e} )B^{\hat{a}\hat{b}}\right) , \\
\Psi &=&\int \sqrt{|\ul{ g} |}%
\bar{C}_{\mu }\left[ \ul{ g}
^{\mu \nu }\ul{ g} ^{\rho
\sigma }\left( D_{\rho }(\ul{ g} )h_{\sigma \nu }+\zeta _{4}D_{\nu }(\ul{ g%
} )h_{\rho \sigma }\right) +\frac{\zeta _{5}}{2}%
\ul{ g} ^{\mu \nu }B_{\nu }\right] ,
\end{eqnarray*}%
in the case of Yang-Mills symmetry (with a simple group, for simplicity),
local Lorentz symmetry and diffeomorphisms, respectively, where $\zeta _{i}$
are constants, $\ul{ A} _{\mu
}^{a}$, $\ul{ e} _{\mu }^{\hat{a%
}}$ and $\ul{ g} _{\mu \nu }$
are the background gauge field, vielbein and metric, $A_{\mu }^{a}$, $f_{\mu
}^{\hat{a}}$ and $h_{\mu \nu }$ are the respective quantum fluctuations and $%
D(\ul{ A} ,\ul{%
 g} )$, $D(\ul{ g%
} )$, $D(\ul{ e} )$ denote the covariant derivatives in the background fields.

The three canonical transformations $F_{\text{b}}$, $F_{\text{nm}}$ and $F_{%
\text{gf}}$ can be composed as follows. The first two commute and have a
vanishing propagator, because the fields (sources) that appear nontrivially
in $F_{\text{nm}}$ have no source (field) counterpart in the nontrivial
sector of $F_{\text{b}}$. Thus, the composition gives the generating
functional%
\begin{equation*}
(F_{\text{b}}\circ F_{\text{nm}})(\Phi ,\ul{ \Phi %
} ,K^{\prime },\ul{ K}%
 ^{\prime })=\int (\Phi ^{\alpha }-\ul{
\Phi } ^{\alpha })K_{\alpha }^{\prime }+\int %
\ul{ \Phi } ^{\alpha }%
\ul{ K} _{\alpha }^{\prime }-\int 
\mathcal{R}_{\bar{C}}^{I}(\bar{C},\ul{ C}%
 )K_{B}^{I\hspace{0.01in}\prime },
\end{equation*}%
and $F_{\text{b}}\circ F_{\text{nm}}=F_{\text{nm}}\circ F_{\text{b}}$.

Now we compose $F_{\text{nm}}$ with $F_{\text{gf}}$. We can consider either $%
F_{\text{nm}}\circ F_{\text{gf}}$ or $F_{\text{gf}}\circ F_{\text{nm}}$.
Applying formula (\ref{fint}), we see that in the first case there is no
nontrivial diagram, since the nontrivial part of $F_{\text{gf}}$ does not
contain sources. Then formula (\ref{c2}) reduces to $C=A+B$ and we obtain%
\begin{equation*}
(F_{\text{nm}}\circ F_{\text{gf}})(\Phi ,\ul{ \Phi %
} ,K^{\prime },\ul{ K}%
 ^{\prime })=\int \Phi ^{\alpha }K_{\alpha }^{\prime }+\int \ul{ \Phi } ^{\alpha }%
\ul{ K} _{\alpha }^{\prime }-\int 
\mathcal{R}_{\bar{C}}^{I}(\bar{C},\ul{ C}%
 )K_{B}^{I\hspace{0.01in}\prime }-\Psi (\Phi ,\ul{%
 \phi } ).
\end{equation*}%
Instead, when we consider $F_{\text{gf}}\circ F_{\text{nm}}$, we have one
nontrivial diagram and formula (\ref{c2}) effectively reduces to $%
C=A+B+A_{i}B^{i}$. Note that the only nontrivial propagator is $(%
\overleftarrow{\delta }/\delta K_{B}^{\prime })(\overrightarrow{\delta }%
/\delta B)$. The composed transformation is%
\begin{equation}
(F_{\text{gf}}\circ F_{\text{nm}})(\Phi ,\ul{ \Phi %
} ,K^{\prime },\ul{ K}%
 ^{\prime })=(F_{\text{nm}}\circ F_{\text{gf}})(\Phi ,%
\ul{ \Phi } ,K^{\prime },%
\ul{ K} ^{\prime })+\int \bar{C}%
^{I}\zeta _{IJ}(\ul{ \phi }
,\partial )\mathcal{R}_{\bar{C}}^{J}(\bar{C},\ul{ C%
} ).  \label{fgfnm}
\end{equation}%
This result can also be found by applying the BCH\ formula (\ref{boso}) for
the composition of the componential maps, with the Poisson brackets replaced
by the antiparentheses (\ref{antip}). We find%
\begin{equation*}
(F_{\text{gf}}\circ F_{\text{nm}})(\Phi ,\ul{ \Phi %
} ,K^{\prime },\ul{ K}%
 ^{\prime })=\mathcal{C}\left( -\Psi (\Phi ,\ul{%
 \phi } )-\int \mathcal{R}_{\bar{C}}^{I}(\bar{C}%
,\ul{ C} )K_{B}^{I\hspace{0.01in%
}\prime }+\frac{1}{2}\int \bar{C}^{I}\zeta _{IJ}(\ul{ \phi } ,\partial )\mathcal{R}_{\bar{C}}^{J}(\bar{C},%
\ul{ C} )\right) .
\end{equation*}%
It is easy to check that only the first two diagrams of (\ref{component})
contribute, so formula (\ref{laffo}) reduces to $\mathcal{C}%
(X)=I+X+(1/2)X_{i}X^{i}$, which gives (\ref{fgfnm}).

In ref. \cite{back} the tensor operator $\zeta _{IJ}$ was set to zero, to
make $F_{\text{gf}}$ and $F_{\text{nm}}$ commute. However, in some
applications, such as the chiral dimensional regularization of ref. \cite%
{chiraldimreg}, which is useful to treat nonrenormalizable general chiral
gauge theories, it is necessary to keep $\zeta _{IJ}$ nonvanishing, to have
well-behaved regularized propagators.

The gauge fixing is the last step of the construction of the action. Indeed,
only after properly organizing the background transformations, it makes
sense to talk about a background invariant gauge fermion. Thus, we must take 
$F_{\text{gf}}\circ F_{\text{nm}}$, rather than $F_{\text{nm}}\circ F_{\text{%
gf}}$.

The composition $F_{\text{gf}}\circ F_{\text{nm}}\circ F_{\text{b}}$ can be
easily worked out by means of formula (\ref{linear}) and gives%
\begin{eqnarray*}
F_{\text{gf}}\circ F_{\text{nm}}\circ F_{\text{b}} &=&\int (\Phi ^{\alpha }-%
\ul{ \Phi } ^{\alpha
})K_{\alpha }^{\prime }+\int \ul{ \Phi }%
 ^{\alpha }\ul{ K}
_{\alpha }^{\prime }-\int \mathcal{R}_{\bar{C}}^{I}(\bar{C},%
\ul{ C} )K_{B}^{I\hspace{0.01in}\prime }
\\
&&-\Psi (\Phi -\ul{ \Phi } ,%
\ul{ \phi } )+\int \bar{C}%
^{I}\zeta _{IJ}(\ul{ \phi }
,\partial )\mathcal{R}_{\bar{C}}^{J}(\bar{C},\ul{ C%
} ).
\end{eqnarray*}%
Applying the composed transformation to the action (\ref{sfb}), we obtain
the background field gauge-fixed action%
\begin{equation*}
S_{\text{b}}=(F_{\text{gf}}\circ F_{\text{nm}}\circ F_{\text{b}})S\text{.}
\end{equation*}

For various applications, it is useful to compare the results of the
background field method with those of the standard, nonbackground approach.
The nonbackground gauge fixed action is $\bar{S}_{\text{nb}}=F_{\text{gf}%
}^{\prime }S$, where%
\begin{equation*}
F_{\text{gf}}^{\prime }(\Phi ,\ul{ \Phi }%
 ,K^{\prime },\ul{ K}
^{\prime })=\int \ \Phi ^{\alpha }K_{\alpha }^{\prime }+\int \ %
\ul{ \Phi } ^{\alpha }%
\ul{ K} _{\alpha }^{\prime }-\Psi
^{\prime }(\Phi )=\mathcal{C}(-\Psi ^{\prime }(\Phi ))
\end{equation*}%
is the generating functional of the canonical transformation that performs
the gauge fixing. The background fields and sources are inert here. As
usual, to simplify the renormalization, it is convenient to take a quadratic
gauge fermion $\Psi ^{\prime }$. We choose%
\begin{equation*}
\Psi ^{\prime }(\Phi )=\int \bar{C}^{I}\left( G^{Ii}(0,\partial )\phi
^{i}+\zeta _{IJ}(0,\partial )B^{J}\right) .
\end{equation*}

For convenience, we further make an irrelevant background shift by applying $%
F_{\text{b}}$, that is to say redefine the nonbackground action as $S_{\text{%
nb}}=(F_{\text{b}}\circ F_{\text{gf}}^{\prime })S$. Then the relation
between the background and nonbackground actions reads%
\begin{equation*}
S_{\text{b}}=(F_{\text{gf}}\circ F_{\text{nm}}\circ F_{\text{b}}\circ F_{%
\text{gf}}^{\prime \hspace{0.01in}-1}\circ F_{\text{b}}^{-1})S_{\text{nb}}.
\end{equation*}%
Formulas (\ref{linear}) and (\ref{linear2}) give%
\begin{equation*}
F_{\text{b}}\circ F_{\text{gf}}^{\prime \hspace{0.01in}-1}\circ F_{\text{b}%
}^{-1}=\int \ \Phi ^{\alpha }K_{\alpha }^{\prime }+\int \ %
\ul{ \Phi } ^{\alpha }%
\ul{ K} _{\alpha }^{\prime }+\Psi
^{\prime }(\Phi +\ul{ \Phi } ).
\end{equation*}%
Using (\ref{fgfnm}) and (\ref{linear}) again, we easily find%
\begin{eqnarray*}
F_{\text{gf}}\circ F_{\text{nm}}\circ F_{\text{b}}\circ F_{\text{gf}%
}^{\prime \hspace{0.01in}-1}\circ F_{\text{b}}^{-1} &=&\int \ \Phi ^{\alpha
}K_{\alpha }^{\prime }+\int \ \ul{ \Phi }%
 ^{\alpha }\ul{ K}
_{\alpha }^{\prime }-\Delta \Psi (\Phi ,\ul{ \Phi %
} )-\int \mathcal{R}_{\bar{C}}^{I}(\bar{C},%
\ul{ C} )K_{B}^{I\hspace{0.01in}\prime }
\\
&&+\int \bar{C}^{I}\zeta _{IJ}(\ul{ \phi %
} ,\partial )\mathcal{R}_{\bar{C}}^{J}(\bar{C},\ul{%
 C} ),
\end{eqnarray*}%
where 
\begin{equation}
\Delta \Psi (\Phi ,\ul{ \Phi }
)=\int \bar{C}^{I}\left( G^{Ii}(\ul{ \phi } ,\partial )\phi ^{i}-G^{Ii}(0,\partial )(\phi ^{i}+%
\ul{ \phi } ^{i})+(\zeta _{IJ}(%
\ul{ \phi } ,\partial )-\zeta
_{IJ}(0,\partial ))B^{J}\right)  \label{deltapsi}
\end{equation}%
is the difference between the background field gauge fermion and the
nonbackground one.

Using the componential map, we find%
\begin{equation*}
F_{\text{gf}}\circ F_{\text{nm}}\circ F_{\text{b}}\circ F_{\text{gf}%
}^{\prime \hspace{0.01in}-1}\circ F_{\text{b}}^{-1}=\mathcal{C}(X),
\end{equation*}%
where%
\begin{equation*}
X=-\Delta \Psi (\Phi ,\ul{ \Phi } )-\int \mathcal{R}_{\bar{C}}^{I}(\bar{C},\ul{ C%
} )K_{B}^{I\hspace{0.01in}\prime }+\frac{1}{2}\int \bar{C}%
^{I}\left( \zeta _{IJ}(\ul{ \phi } ,\partial )+\zeta _{IJ}(0,\partial )\right) \mathcal{R}_{\bar{C}}^{J}(%
\bar{C},\ul{ C} ).
\end{equation*}%
Again, formula (\ref{laffo}) reduces to $\mathcal{C}(X)=I+X+(1/2)X_{i}X^{i}$%
, because the only nontrivial propagator is $(\overleftarrow{\delta }/\delta
K_{B}^{\prime })(\overrightarrow{\delta }/\delta B)$ and $X$ is linear in $B$%
, $K_{B}^{\prime }$.

We can continuously interpolate between the background and nonbackground
approaches by introducing a parameter $\xi $ that varies from $0$ to $1$ and
considering the canonical transformation generated by%
\begin{equation}
F_{\xi }=\mathcal{C}(\xi X).  \label{cix}
\end{equation}%
Explicitly, we find 
\begin{eqnarray}
F_{\xi }(\Phi ,\ul{\Phi },K^{\prime },\ul{K}^{\prime
},\xi ) &=&\int \ \Phi ^{\alpha }K_{\alpha }^{\prime }+\int \ %
\ul{\Phi }^{\alpha }\ul{%
K}_{\alpha }^{\prime }-\xi \Delta \Psi -\xi
\int \mathcal{R}_{\bar{C}}^{I}(\bar{C},\ul{C})K_{B}^{I\hspace{0.01in}\prime }  \notag \\
&&+\frac{\xi }{2}\int \bar{C}^{I}\left[ (1+\xi )\zeta _{IJ}(%
\ul{\phi },\partial )+(1-\xi )\zeta
_{IJ}(0,\partial )\right] \mathcal{R}_{\bar{C}}^{J}(\bar{C},%
\ul{C}).  \label{passo}
\end{eqnarray}

Note that the h-logarithm of (\ref{cix}) is equal to $X$ with $K_{B}^{I%
\hspace{0.01in}\prime }$ replaced by $K_{B}^{I}$ and plays the role of the $%
\xi $-independent Hamiltonian.

A different interpolation amounts to taking, for example,%
\begin{equation}
F_{\xi }^{\prime }=\int \ \Phi ^{\alpha }K_{\alpha }^{\prime }+\int \ \ul{ \Phi } ^{\alpha }%
\ul{ K} _{\alpha }^{\prime }-\xi \Delta
\Psi (\Phi ,\ul{ \Phi } )-\xi
\int \mathcal{R}_{\bar{C}}^{I}(\bar{C},\ul{ C} )K_{B}^{I\hspace{0.01in}\prime }+\xi \int \bar{C}^{I}\zeta
_{IJ}(\ul{ \phi } ,\partial )%
\mathcal{R}_{\bar{C}}^{J}(\bar{C},\ul{ C}%
 ).  \label{p2}
\end{equation}%
The h-logarithm of this expression gives a $\xi $-dependent Hamiltonian,
which we now calculate.

Assume that $U(\Phi ,K,\xi )$ is a function that behaves as a scalar under
canonical transformations $\Phi ,K\rightarrow \Phi ^{\prime },K^{\prime }$,
i.e. such that $U^{\prime }(\Phi ^{\prime },K^{\prime },\xi )=U(\Phi
,K,\xi )$. Then formula (\ref{remo}) turns into \cite{removal} (see also
the appendix of \cite{back})%
\begin{equation}
\frac{\partial U^{\prime }}{\partial \xi }=\frac{\partial U}{\partial \xi }%
-(U,Y),\qquad Y(\Phi ,K,\xi )=\widetilde{\frac{\partial \mathcal{F}}{%
\partial \xi }},  \label{removal}
\end{equation}%
where $\mathcal{F}(\Phi ,K^{\prime },\xi )$ is the generating functional of
the canonical transformation and the tilde means that, after taking the $\xi 
$ derivative, the source $K^{\prime }$ must be expressed in terms of $\Phi $%
, $K$ and $\xi $. Choosing $\mathcal{F}=F_{\xi }^{\prime }$ and enlarging
the sets of fields and sources to include the background ones, we find the
h-logarithm 
\begin{equation*}
Y(\Phi ,K,\ul{\Phi },%
\ul{K},\xi )=-\Delta \Psi (\Phi ,\ul{\Phi })-\int \mathcal{R}_{\bar{C}%
}^{I}(\bar{C},\ul{C})K_{B}^{I}+\int \bar{C}^{J}\left[ (1-\xi )\zeta _{JI}(\ul{%
\phi },\partial )+\xi \zeta _{JI}(0,\partial )%
\right] \mathcal{R}_{\bar{C}}^{I}(\bar{C},\ul{C%
}).
\end{equation*}%
It may be more convenient to work with the interpolation (\ref{passo}),
whose h-logarithm is $\xi $ independent, rather than (\ref{p2}).

The dependence of the correlation functions on the parameters introduced by
a canonical transformation is encoded into the equations of gauge dependence 
\cite{gaugedep,voronov,removal,kraus,recentgaugedep}, sometimes known as
Nielsen identities. The componential map and the other tools of this paper
may be convenient to manipulate those equations more efficiently. In
particular, the interpolation (\ref{cix}) allows us to take advantage of
the background field method and prove key properties of renormalization in
simpler, more powerful ways. An illustration of this fact can be found in
ref. \cite{KSZ}, where an important theorem about the cohomology of
renormalization was proved. That theorem allows us to classify the
structures of the counterterms and the local contributions to anomalies.

In turn, the classification of counterterms and anomalies is important to
show, to all orders of the perturbative expansion, that the gauge
symmetries are not affected by the subtraction of divergences (up to
canonical transformations). The background field method and the
interpolation (\ref{p2}) have been used \cite{back} to achieve this goal in
manifestly nonanomalous theories, renormalizable or not. In potentially
anomalous nonrenormalizable theories, such as the standard model coupled to
quantum gravity, which require a more involved regularization \cite%
{chiraldimreg}, the goal must be achieved together with the proof of the
Adler-Bardeen theorem \cite{adlerbardeen,ABnonreno} for the cancelation of
anomalies to all orders (when they vanish at one loop). Within the standard,
nonbackground approach, this was done for the first time in ref. \cite%
{ABnonreno}. The techniques of this paper and the results of \cite{KSZ} may
be useful to upgrade the derivation of \cite{ABnonreno} to the background
field approach and prepare the ground to make further progress.

\section{Conclusions}

\setcounter{equation}{0}

\label{s7}

Canonical transformations play an important role not only in classical
mechanics, but also in quantum field theory. In several situations, it is
useful to have practical formulas for the perturbative expansion of the
generating functions around the identity map. In this paper we have given a
number of such formulas, starting from the composition law, which we have
expressed as the tree sector of a functional integral and later rephrased by
means of the componential\ map.

The componential map is a standard way to express the generating function of
a canonical transformation. It makes the inverse operation straightforward
and obeys the Baker-Campbell-Hausdorff\ formula. It also admits a simple
diagrammatic interpretation and a time-ordered generalization. It can be
related to the solution of the Hamilton-Jacobi equation, expressed as a
perturbative expansion in powers of a suitable Hamiltonian, its derivatives
and its integrals over time.

The formulas we have found can be straightforwardly generalized from
classical mechanics to quantum field theory, where the functionals and the
conjugate variables may have both bosonic and fermionic statistics.
Particularly interesting are the applications to the Batalin-Vilkovisky
formalism. Canonical transformations are commonly used to implement the
gauge fixing, make arbitrary changes of field variables and changes of the
gauge fixing itself, switch to the background field method and subtract the
counterterms proportional to the field equations. Various times these
operations must be composed and inverted. Practical formulas, such as the
ones given in this paper, allow us to handle these operations quickly. In
particular, they can be convenient in nonrenormalizable theories, where the
cohomology of counterterms and anomalies involves nonpolynomial functionals
and the renormalization of divergences involves nonpolynomial canonical
transformations.

\end{document}